\documentclass[aps,pre,amssymb,amsmath]{revtex4}
\usepackage{floatflt}
\usepackage[dvips]{graphicx}
\usepackage{hhline}

\newcommand{\be}{\begin{equation}}
\newcommand{\ee}{\end{equation}}
\newcommand{\bea}{\begin{eqnarray}}
\newcommand{\eea}{\end{eqnarray}}
\newcommand{\p}{\partial}
\newcommand{\vvect}{\mathbf{v}}
\newcommand{\mydiv}{\mathrm{div}\,}
\newcommand{\rot}{\mathrm{rot}\,}
\begin{document}
\title{Evaporation and fluid dynamics
of a sessile drop of capillary size}
\author{L.Yu.\ Barash\,$^{1,a)}$, T.P.\ Bigioni\,$^{2}$,
V.M.\ Vinokur\,$^{3}$, L.N.\ Shchur\,$^{1,3}$}
\affiliation{
$^{1)}$ Landau Institute for Theoretical Physics, 142432
Chernogolovka, Russia \\
$^{2)}$ James Franck Institute, The University of Chicago,
Chicago, Illinois 60637, USA  \\
$^{3)}$ Materials Science Division, Argonne National Laboratory,
Argonne, Illinois 60439, USA \\
e-mail: \tt $^{a)}$barash@itp.ac.ru}
\begin{abstract}

Theoretical description and numerical simulation of an evaporating
sessile drop are developed. We jointly take into account the
hydrodynamics of an evaporating sessile drop,
effects of the thermal conduction in the drop and
the diffusion of vapor in air.
A shape of the rotationally symmetric drop is determined
within the quasistationary approximation.
Nonstationary effects in the diffusion
of the vapor are also taken into account.
Simulation results agree well with the data of
evaporation rate measurements for the toluene drop.
Marangoni forces associated with the temperature
dependence of the surface tension, generate fluid convection
in the sessile drop.
Our results demonstrate several dynamical stages of the convection
characterized by different number of vortices in the drop.
During the early stage the street of vortices arises near
a surface of the drop and induces a non-monotonic spatial
distribution of the temperature over the drop surface.
The initial number of near-surface vortices in the drop
is controlled by the Marangoni cell size which is similar
to that given by Pearson for flat fluid layers.
This number quickly decreases with time,
resulting in three bulk vortices in the intermediate stage.
The vortices finally transform into the single convection vortex
in the drop, existing during about $1/2$ of the evaporation time.
\end{abstract}
\maketitle

\section{Introduction}

Evaporation of a liquid drop in an ambient gas is a well-known yet not completely solved problem of the classical physics (see, for example,~\cite{Langmuir1918,Fuchs}).  Past decade was marked by significant advance in experiment and progress in understanding several key aspects of evaporation process, in particular, the
vapor diffusion from the sessile drop surface and the
hydrodynamic effects within the evaporating drops~\cite{Deegan97,Deegan,
HuLarsonEvap,Popov1,HuLarsonMarangoni,Girard,Ristenpart}. It was found, in particular, that the evaporating flux density is inhomogeneous along the surface and diverges on approach to the pinned contact line~\cite{Deegan97,Deegan}.  The resulting inhomogeneous mass flow modifies the
temperature distribution over the drop surface and, hence, the
Marangoni forces associated with the temperature dependent surface
tension. The convection inside a droplet~\cite{Zhang,Davis1,Davis2,
Davis3,Davis4,Lozinsky,Niazmand,Rednikov,Savino,HuLarsonMarangoni,
Girard,Xu} appears quite different from the classical
Marangoni convection in the systems with a simple flat geometry~\cite{Benard,
Pearson}. Thermal conductivity of the substrate can also influence
the formation of flows within a liquid drop since it is the magnitude of the conductivity which determines
the sign of the tangential component of the temperature gradient at the
surface close to the contact line and, therefore, the direction of
the convection~\cite{Ristenpart}. The observation of the distinct stages
of the evaporation process~\cite{PicknettBexon,Shanahan1,Shanahan2}
has revealed that the longest and dominating regime of the evaporation process is the constant contact
area mode, where the contact line is pinned.  As further the contact line gets depinned, the different regime, the constant contact
angle mode switches on. Finally,
the drying mode follows, in which the height, the contact area and the contact angle
rapidly decrease with time.

Understanding the details of the drop dynamics and evaporation is crucial to many applications involving this process, including preparing ultra-clean
surfaces~\cite{Leenaars90,Marra91,Huethorst91,Matar01}, protein crystallography~\cite{Denkov,Dimitrov},
the studies of DNA stretching behavior and DNA mapping
methods~\cite{Jing,HuLarsonDNA,Hsieh}, developing methods for jet ink printing~\cite{Park,Jong,Lim},
and in many other fields (see, for example,~\cite{Frohn}).
The process of evaporation of the drop with the colloidal suspension in it is of interest for
methods of fabrication of various structures
on the substrate.
One of the examples is the effect
of evaporative contact line deposition, the
so-called coffee-ring effect~\cite{Deegan97,Deegan,Govor,HuLarsonCoffee,
Popov1,Popov2}. Other important example is
the self-assembly of long-range-ordered nanocrystal
superlattice monolayer~\cite{Lin1,Lin2,Bigioni3}.

While the fundamentals of quasistationary evaporation behavior 
are established, to some extent~\cite{HuLarsonMarangoni,Girard}, 
the full dynamical description of the liquid drop evaporation is yet not available.  In this work we undertake a step towards such a description offering a quantitative approach which enables us to account for all the relevant components of the evaporation process,  namely, the fluid dynamics, the vapor diffusion, and the spatial temperature distribution in an
evaporating sessile drop.  The calculation are carried out according to the following scheme: (i) We derive the equation yielding the shape of the sessile drop and find how the evaporation rate is influenced by the deviations of the drop shape from the ideal spherical cap caused by the gravitational force. (ii) We solve the diffusion equation describing the vapor kinetics and calculate the local evaporation flux from the surface of the drop as well as the resulting evolution of the drop evaporation rate with time.  At this stage we take into account the nonstationary effects in a
vapor diffusion and find that the dynamical corrections to the
evaporation rate do not vanish exponentially, but decay as $\propto 1/\sqrt{t}$.
(iii) We solve the thermal conduction equation and find
the spatial temperature distribution in the drop.
(iv) We solve the Navier-Stokes equations and derive the
velocity field corresponding to the convection within the drop.
(v)  Finally we arrive at a full description of the time evolution of the temperature and the fluid
convection in the drop and identify characteristic stages
of the thermocapillary convection.  To this end we take into account inertial terms in the
Navier-Stokes equations, including the time derivatives.

To ensure self consistency of the derivation of the physical characteristics of the system we include the temperature variation over the drop surface (since the surface tension depends on temperature) into the boundary conditions for the fluid dynamics in the drop.  Furthermore, we take into account that the velocity field
influences the thermal conduction, since the velocities enter
the thermal conduction equation. And, finally, we take into account that the local evaporation flux
is related to the heat transfer and, hence, to the temperature gradient
on the drop surface.

While the developed approach is quite general and thus applies to a wide variety of the evaporating problems, we focus here on the experimental situation corresponding to evaporation of the toluene sessile drop of the capillary size. The typical values of the characteristic parameters at $T_0=295$K are taken from~\cite{Lide} and are presented in the lower part of Table~\ref{paramtable}. The initial values which we use in our calculations are given in the upper
part of Table~\ref{paramtable} and correspond to the evaporating toluene drop
containing the colloidal solution of gold
nanoparticles.  These nanoparticles stick to the drop surface making islands there and form a monolayer nanocrystal on the substrate after toluene dries out~\cite{Lin1,Lin2,Bigioni3}.

\begin{table}[htb]
\caption{ The parameter values used in the paper for
modeling the evaporation and hydrodynamics of the drop.
The tabular data are taken at $T_0=295$K from~\cite{Lide}.}
\begin{tabular}{|c|l|l|}
\hline
Drop & Contact line radius             & $r_0=0.2$ cm \\
parameters & Initial height             & $h=0.1314$ cm\\
&Initial mass               & $m_0=8.7\cdot 10^{-3}$ g \\
\hline
\hline
Toluene
&Density                & $\rho=0.87$ g/cm$^3$ \\
characteristics
& Molar mass & $\mu=92.14$ g/mole\\
&Thermal conductivity   & $k=1.311\cdot 10^{-3}$ W/(cm$\cdot$ K) \\
&Thermal diffusivity    & $\kappa=k/(\rho c_p)=8.86\cdot 10^{-4}$ cm$^2$/s \\
&Kinematic viscosity    & $\nu=6.4\cdot 10^{-3}$\ cm$^2$/s \\
&Dynamic viscosity      & $\eta=\nu\rho=5.6\cdot 10^{-3}$ g/(cm$\cdot$ s) \\
&Specific heat          & $c_V=1.51$\ J/(g$\cdot$K) \\
&Thermal-expansion coefficient   & $\beta=1.138\cdot 10^{-3}$\ K$^{-1}$ \\
&Surface tension        & $\sigma=28.3049$ g/s$^2$ \\
&Temperature derivative of surface tension  & $\sigma'=-\p\sigma/\p T=0.1189$\ g/(s$^2\cdot$ K) \\
&Capillary constant      & $a=0.258$ cm \\
&Latent heat of evaporation  & $L=300$ J/g \\
\hline
\hline
Toluene
&Diffusion constant & $D=0.1449$ cm$^2$/s\\
vapor
&Saturated toluene vapor density & $u_s=1.27\cdot 10^{-4}$ g/cm$^3$ \\
characteristics&Mean free path & $\lambda=2.7\cdot 10^{-5}$ cm \\
\hline
\hline
Air
&Thermal conductivity          & $k_{a}=0.258\cdot 10^{-3}$ W/(cm$\cdot$ K)\\
characteristics
&Dynamical viscosity     & $\eta_{a}=1.82\cdot 10^{-4}$ g/(cm$\cdot$ s) \\
&Density                 & $\rho_{a}=1.2\cdot 10^{-3} $ g/cm$^3$ \\
&Kinematic viscosity   & $\nu_{a}=\eta_{a}/\rho_{a}=0.15$ cm$^2$/s\\
\hline
\end{tabular}
\label{paramtable}
\end{table}

We have identified three major distinct dynamic stages of the Marangoni convection during the evaporation of the toluene drop.  During the initial stage, vortices appear near the surface of the drop, during the second stage they grow, coalesce and rapidly migrate into the drop bulk, and, finally, a single vortex survives governing all the fluid dynamics of the thermocapillary
effect in the system.
We derive the temperature profiles, evaporation rates,
and the velocity distributions for each stage of the convection and uncover the important role of the convective heat transfer and the effect of inertial terms in the
Navier-Stokes equations in the evaporation dynamics.

The paper is organized as follows.
In Sec.~\ref{LargeSection} we
present basic equations and main analytical results along with the quantitative description of main dynamical processes.
In particular, Sec.~\ref{DropShapeSec} deals with the shape of a sessile drop.
Various aspects of the vapor
diffusion and the drop evaporation rates
are discussed in Sec.~\ref{DiffusionSec}--\ref{SessileSpherical}.
The characteristic Marangoni numbers
for the drop are found in Sec.~\ref{MarangoniSec}.
Sec.~\ref{EqSec} makes use of a self-consistent approach for calculating
the velocity field and the spatial temperature.
In Sec.~\ref{ResultsSec}
we present our main results and discussion, including
the experimental and theoretical data for the evaporation
rates and contact angles, the dynamics of vortex structure
and the temperature profile. Sec.~\ref{DiscussionSec} contains discussion.
The details of the developed numerical approach are described
in Appendix~\ref{NumericalSec}.

\section{Basic equations and formulas}
\label{LargeSection}

\subsection{The sessile drop shape}
\label{DropShapeSec}

During the evaporation process a drop loses its mass, hence
with time its volume decreases and the shape changes.
In this section we present the quasistationary
method for calculating the shape of a sessile
drop based on the hydrostatic Laplace equation.
The approach holds as long as viscous forces,
which generally enter the boundary condition for the
pressure, are small. The ratio of
viscous to capillary forces is characterized by the dimensionless
number $C_a=\eta \overline{v}/\sigma$, where $\overline{v}$
is the characteristic value of the velocity.
In our case $C_a\approx 2\cdot 10^{-4}\ll 1$.

The Laplace equation states that the pressure
difference $\Delta p$, taken at the different sides
of the surface of the liquid in an arbitrary point equals  $\sigma k$, where
$k=1/{R_1}+1/{R_2}$
is the mean curvature of the surface and $\sigma$ is the
surface tension~\cite{LL6}. Taking into account the gravity, one
finds that at the surface of the drop
\be
k+\frac{\rho gz}\sigma=k+\frac{2f}{a^2}=const,
\label{StaticLaplace}
\ee
where the shape of the drop surface is defined by the relation $z=f(x,y)$  and
$a=\sqrt{{2\sigma}/{(\rho g)}}$ is the
capillary constant.  The curvature of the surface of the drop is, in its turn, also expressed in terms
of the function $f(x,y)$. Indeed, $k=-\mbox{Tr}(G^{-1}Q)$, where
$G$ and $Q$ are matrices of the first and the second quadratic
forms of the surface~\cite{Dubrovin}. Plugging in the surface equation $z=f(x,y)$
we find
\be
G=\left(\begin{array}{cc} 1+f_x^2&f_xf_y\\f_xf_y&1+f_y^2\\
\end{array}\right),
\qquad\qquad
 Q=\frac1{\sqrt{1+f_x^2+f_y^2}} \left(\begin{array}{cc}
f_{xx}&f_{xy}\\f_{xy}&f_{yy}\\ \end{array}\right).
\ee

\noindent Therefore,

   \be
        k=-\mbox{Tr}(G^{-1}Q)
        =\frac{2f_xf_yf_{xy}-f_{xx}(1+f_y^2)-f_{yy}(1+
        f_x^2)}{(1+f_x^2+f_y^2)^{3/2}}.
        \label{k-3d}
     \ee
We consider a sessile drop with the axial symmetry, where $z=f(r)$,
$r=\sqrt{x^2+y^2}$.  Then Eq.~(\ref{k-3d}) assumes the form
   \be
     k=\frac{f''}{(1+f'^2)^{3/2}}+\frac{f'}{r(1+f'^2)^{1/2}},
   \ee

\noindent where prime denotes the derivative with respect to $r$.
It is convenient to parametrize and separate the variables as~\cite{BA1883,RBN}:
  \be
    \label{rz-separation}
    r=r(s),\qquad z=z(s),
  \ee
where $s$ is the arc length on the drop surface taken from the apex.
The curvature radii are $R_1$ and $R_2$, where $R_1$
is the curvature radius in the meridian $rz$-plane,
and $R_2$ is the second curvature radius.
Both curvature radii can be expressed in terms of
the angle $\phi$ between the normal vector to the drop surface
and the symmetry axis. Namely,
    \be
         \frac1{R_1}=\frac{d\phi}{ds},\qquad R_2=\frac{r}{\sin\phi}.
    \ee
Also, $\p r/\p s=\cos\phi$, $\p z/\p s=-\sin\phi$
and $r(0)=z(0)=\phi(0)=0$.
Introducing the curvature radius $R_0$ at the apex,
we can rewrite the Laplace equation as
   \be \label{RotSymLaplace}
    \sigma\left(\frac{d\phi}{ds}+\frac{\sin\phi}{r}\right)=
    \frac{2\sigma}{R_0}-\rho gz
  \ee
and therefore
   \be
      \frac{d\phi}{ds}=\frac2{R_0}-\frac{\rho gz}{\sigma}-\frac{\sin\phi}{r}.
   \ee
\noindent Hence, for
the column vector $\mathbf{y}=(r(s),\phi(s),z(s))^T$,
where $T$ is the transposition operation,
we have the set of first-order differential
equations with Cauchy boundary conditions:
   \bea
         \label{Cauchy1}
     \frac{d\mathbf{y}}{ds}&=&\mathbf{f}(s,\mathbf{y}),\\
         \label{Cauchy2}
     \mathbf{f}(s,\mathbf{y})&=&\left(\cos\phi,
      \frac2{R_0}-\frac{\rho gz}{\sigma}-\frac{\sin\phi}{r},-\sin\phi\right)^T,\\
        \label{Cauchy3}
     \mathbf{y}(0)&=&\left(0,0,0\right)^T,\qquad
     \mathbf{f}(0,\mathbf{y})=\left(1,\frac1{R_0},0\right)^T,
   \eea
where the last relationship ensures that
$\phi(s)/r(s)\to 1/R_0$ for $s\to 0$.  The Cauchy problem (\ref{Cauchy1})-(\ref{Cauchy3}) can be solved
using the fourth order Runge-Kutta method or the
Adams-Bashforth method~\cite{BA1883}. Therefore, given the values
of $R_0$ and $s_{max}$, one finds the drop shape in the form of
$\mathbf{y}(s)=(r(s),\phi(s),z(s))^T$.

The quantities that are measured in the experiment are the mass of the drop
   \be
     \label{mass}
     m=\pi\rho\int_0^{s_{max}} r^2(s)\sin\phi(s)ds\,\,\, ,
   \ee
\noindent and the radius $r_0$ of the substrate to which the drop is
pinned. Given the values of $R_0$ and $r_0$,
one can easily find $s_{max}$, and, therefore, $m$.
Inverting the function $m(R_0,r)$ numerically, one can
obtain the value of $R_0$ and a drop shape for a given mass $m$
and contact line radius $r_0$.
We note that Eqs.~(\ref{Cauchy1})-(\ref{Cauchy3}) are exact
for a sessile drop with axial symmetry.

The evaporation rate for a given drop shape can be obtained as
\be
\label{dmdtEq}
\left|\frac{dm}{dt}\right|=\int_0^{s_{max}} 2\pi r(s) J(r(s)) ds,
\ee
where $J(r(s))$ is the mass evaporated per second from
unit area of the surface, i.e. the local evaporation flux.

\subsection{Evaporation of the drop}

\label{DiffusionSec}

Consider a sessile droplet resting on a flat substrate.
The vapor concentration above the droplet is time dependent
and inhomogeneous during the evaporation process.
Since the diffusion of the vapor from a near-surface layer
is slower than the evaporation~\cite{LL6},
the vapor concentration at the surface of the drop is assumed to
equal the saturation value $u_s$.
Far above the drop, the toluene vapor concentration is negligible.
The dynamics of the vapor concentration in the surrounding atmosphere
is described by the diffusion equation

\begin{equation}
\frac{\p u}{\p t}=D\Delta u .
\label{Diffusion}
\end{equation}

\begin{figure}[htb]
\caption{ (Color online.) Local evaporation flux density $J(r,t)$ from
a fixed surface as a function
on $r$ for $t=1,2,3,5,10,50,100,500,1000,4000$
seconds of the evaporation process, from top to bottom, respectively
(black curves).
The time dependence of the vapor concentration is taken into
account, whereas the surface of the drop with $m=8.7$ mg is fixed.
The surface is taken as a spherical cap (left panel)
and as a sessile drop surface (right panel).
The bottom line on the left panel (blue curve)
represents $J(r,\infty)$
taken from the exact solution (\ref{JDeegan}).
The bottom line on the right panel (green curve) 
represents $J(r,\infty)$ for a sessile drop.}
\label{round_drop_evap_versus_time}
\includegraphics[width=0.49\textwidth]{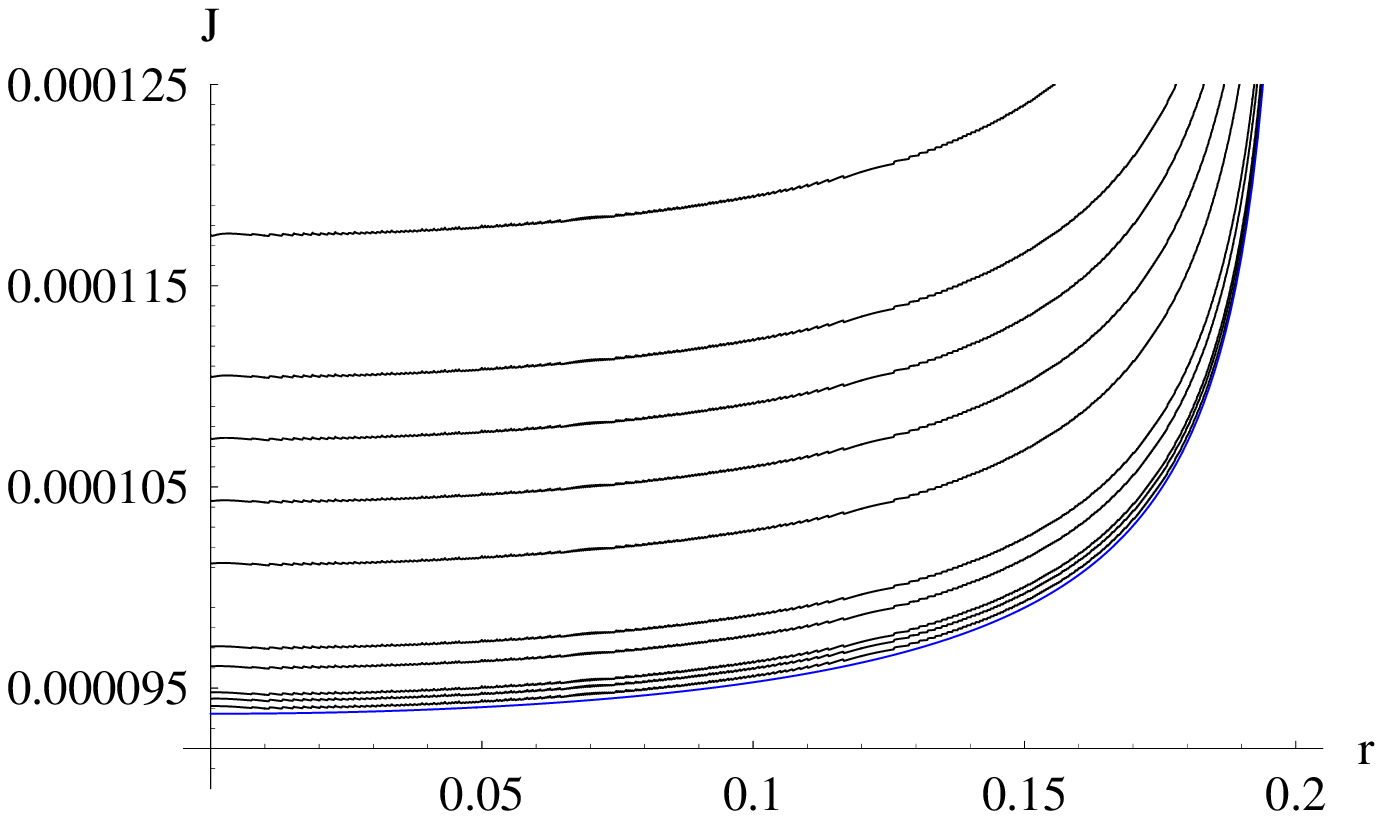}
\includegraphics[width=0.49\textwidth]{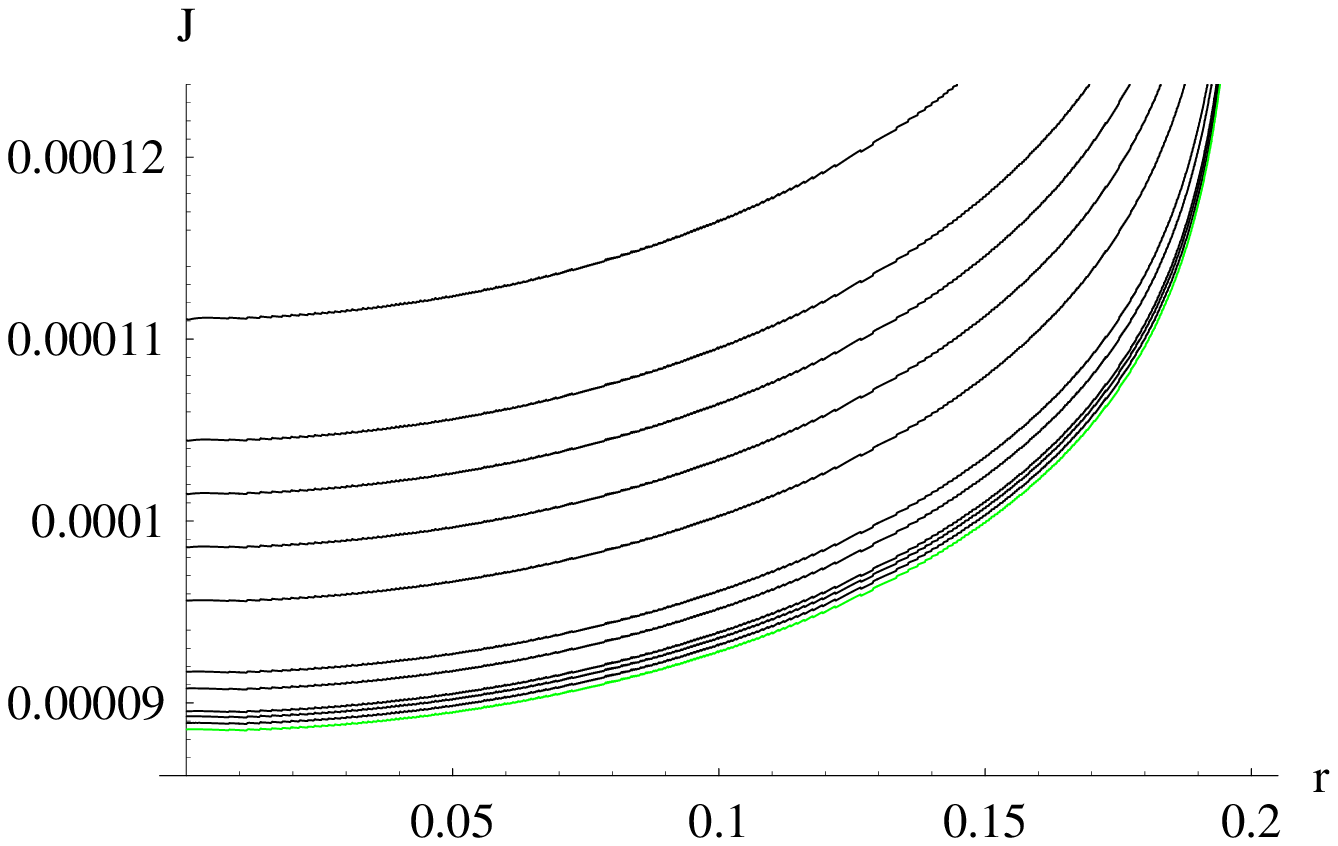}
\end{figure}

We carry out our calculations
(see Sec.~\ref{ResultsSec} and Appendix~\ref{NumericalSec})
taking into account both time dependence of vapor
concentration and the deviations of the sessile drop shape
from a spherical cap. It is instructive to begin
discussing the problem with the more simple case
allowing for an analytical description. To this end we notice that if evaporation can be viewed as an adiabatic process in a sense that the vapor concentration adjusts fast enough to the change of the drop size (and shape) i.e. on the time scales much less than the droplet evaporation
time $t_f$, then the diffusion equation
(\ref{Diffusion}) can be replaced by the Laplace's equation $\Delta u=0$,
and the evaporation kinetics can be considered as a steady state process.  Indeed the time $t_p$ required
for the Brownian particle to pass the characteristic length
$r_0$ is $t_p=r_0^2/D\ll t_f$ under the experimental conditions $t_p\approx 0.2$ s and $t_f\approx 500$ s.
At the same time it interesting as well to consider particular
dependencies of the local evaporation rates on time.
Our simulation results based on Eq.~(\ref{Diffusion})
for the drop with the fixed surface
can be fitted with the
power-law time dependence
   \be
     J(r,t)=J(r,\infty)\left(1+\frac{Ar_0}{2\sqrt{Dt}}\right),
     \label{Jfit}
   \ee
which is almost exact with the accuracy within 1\% for $t>0.5$\,s and $|r_0-r|>0.01$\,cm
(see Fig.~\ref{round_drop_evap_versus_time}).
Here $J(r,t)=|D{\mathbf\nabla} u|$ is the local evaporation flux
on the drop surface, and constant $A = 0.966$ is the only fitting parameter.
For the case of the spherical cap the asymptotic value of the evaporation flux density
$J(r,\infty)$ can be related to $J(\xi,\theta)$ from Eq.~(\ref{JDeegan}).
Eq.~(\ref{Jfit}) and Fig.~\ref{round_drop_evap_versus_time}
confirm the above estimation that the vapor
concentration becomes stationary under
the condition $t \gg r_0^2/D$.
At the same time, for $t=t_f$ the second term in~(\ref{Jfit})
still exceeds $1\%$ of the first term.
The time dependence
$J(r,t)-J(r,\infty)\propto 1/\sqrt{t}$ in Eq.~(\ref{Jfit})
is known for an isotropic diffusion
when the vapor concentration is kept saturated on
the fixed surface of the sphere~\cite{Fuchs}.
This is also valid for a diffusion
from a flat plane~\cite{LL6}.
Our results demonstrate that such dependence also takes
place for an inhomogeneous diffusion from a fixed surface
of the sessile drop.

We further derived the numerical results which take into account the
time dependent drop profile (see Fig.~\ref{fig-dmdt}).
In particular, they show that the corrections for the evaporation rate
due to the nonstationary effects may be up to $5\%$ of
the resulting value.

Deegan et al.\cite{Deegan} reported an analytical solution
for a stationary spatial distribution of the vapor concentration
for a drop with the shape of a spherical cap.
The problem was shown to be mathematically
equivalent to that solved by
Lebedev~\cite{Lebedev}, who obtained the electrostatic potential
of a charged conductor having the shape defined by two
intersecting spheres.  Modifying the equations derived in~\cite{Deegan,Lebedev},
we arrive at the following formula for the vapor concentration $u(\alpha,\beta)$ above
the surface (\ref{SphericalCap}) of a spherical drop:
\begin{equation}
u(\alpha,\beta)=u_\infty+(u_s-u_\infty)\sqrt{2(\cosh\alpha-\cos\beta)}
\int_0^\infty \frac{\cosh\theta\tau}{\cosh\pi\tau}
\frac{\cosh\beta\tau}{\cosh(\pi-\theta)\tau} P_{-1/2+i\tau}(\cosh\alpha)d\tau.
\label{uDeegan}
\end{equation}
Here $\theta$ is the drop contact angle,
$-\pi+\theta \le \beta \le \pi-\theta$,
$u_\infty=0$, $\alpha$ and $\beta$ are the toroidal coordinates:
\begin{equation}
r=\frac{r_0\sinh\alpha}{\cosh\alpha-\cos\beta},\qquad
z=\frac{r_0\sin\beta}{\cosh\alpha-\cos\beta}.
\end{equation}
At the drop surface $\beta=\pi-\theta$, therefore
$\xi=r/r_0={\sqrt{x^2-1}}/{(x+\cos\theta)}$, where
$x(\xi,\theta)=\cosh\alpha$, hence,
\begin{equation}
x(\xi,\theta)=\cosh\alpha=
\frac{\xi^2\cos\theta+\sqrt{1-\xi^2\sin^2\theta}}{1-\xi^2}.
\label{XToroidal}
\end{equation}
The evaporation flux at the surface is
\be
J=|D{\mathbf\nabla} u|=
\frac{D}{r_0}(\cosh\alpha-\cos\beta)
\left.\frac{\p u}{\p\beta}\right|_{\beta=\pi-\theta},
\ee
and, therefore,
\begin{equation}
J(\xi,\theta)=\frac{Du_s}{r_0}\left(\frac{\sin\theta}2+
\sqrt{2}(x(\xi,\theta)+\cos\theta)^{3/2}
\int_0^\infty\frac{\cosh\theta\tau}{\cosh\pi\tau}\tau\tanh(\pi-\theta)\tau
P_{-1/2+i\tau}(x(\xi,\theta)) d\tau\right).
\label{JDeegan}
\end{equation}

\begin{figure}[htb]
\includegraphics[width=0.6\textwidth]{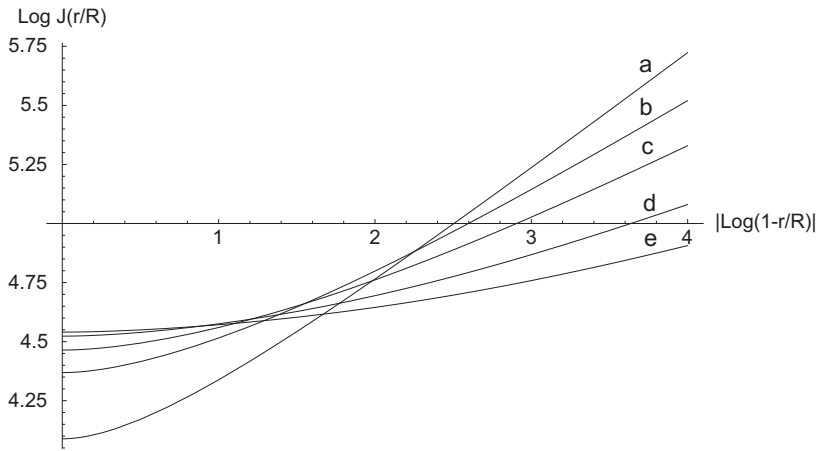}
\caption{Log-log plot of $J(r)$
for drops with different contact angles.
Lines marked with $a,b,c,d,e$ correspond to
values $\pi/120$, $\pi/6$, $\pi/4$, $\pi/3$, $23\pi/60$
of the contact angle $\theta$, respectively.}
\label{logplot}
\end{figure}

After substituting $\xi=r/r_0$ and~(\ref{XToroidal}) in Eq.~(\ref{JDeegan}),
one obtains $J(r)$. Fig.~\ref{logplot} shows $\log J(r)$
for drops with different values of the contact angle $\theta$.

The total drop mass is
\be
m=\frac{\rho\pi r_0^3}{6}\tan\frac{\theta}{2}
\left(3+\tan^2\frac{\theta}{2}\right),
\ee
and the total evaporation rate is
\be
\left|\frac{dm}{dt}\right|=
2\pi r_0^2\int_0^1\frac{\xi J(\xi,\theta)d\xi}{\sqrt{1-\xi^2\sin^2\theta}}.
\label{dmfromdt}
\ee

While it were possible to simplify the expression for the drop evaporation rate
adapting the sessile drop shape to that of the spherical cap
of the same mass (see Sec.~\ref{SessileSpherical}), and then integrate Eqs.(\ref{JDeegan}),~(\ref{dmfromdt}),
we will employ a more general method and
develop a description of the evaporation rate which takes into account
altogether spatial and time variations of the
vapor diffusion and the sessile drop shape.

\subsection{Estimation for the evaporation time}
\label{RoughEstim}

In this section we derive estimates for the evaporation
time of the drop, which in spite of their relative simplicity, offer, nevertheless, a
fairly good description of the experimental results. Let us consider a spherical evaporating drop of radius $r_0=0.2$ cm,
so that the vapor density is saturated at the drop surface
and vanishes far away from the drop.
The diffusion of the toluene vapor controls the evaporation process.
The outward flow of the vapor through the concentric sphere around the drop
of the radius $R$ is $J=-D\cdot 4\pi R^2 {du}/{dR}$
and does not depend on $R$.
Here $u$ is the vapor density on the surface of the
sphere of radius $R$, and $D$ is the diffusion constant
for the toluene vapor.
Therefore, $R^2 du/dR=-A$, where $A=J/(4\pi D)$.
Since the vapor concentration far away from the drop is negligible,
we obtain $u=A/R$, $A=r_0u_s$ and $J=4\pi D r_0u_s$.
On the other hand, the flow of molecules from the fluid drop is
$J=-\rho d(4\pi r_0^3/3)/dt$.
Therefore, $r_0dr_0/dt=-D u_s/\rho$, and
the evaporation time of the drop can be written as
\be
t=\frac{\rho}{u_s}\frac{r_0^2}{2D}.
\label{evap-time}
\ee
Eq.~(\ref{evap-time}) is a well-known result of
the classical Maxwell's theory of the evaporation~(see,
for example,~\cite{Fuchs}).
The diffusion constant for toluene vapor can be estimated as
\be
D=\frac{1}{3} \langle v\rangle \lambda.
\label{diff-const}
\ee
Here $\langle v\rangle=\sqrt{{8RT}/{(\pi\mu)}}=26035$ cm/s is the
average velocity for the thermal motion of the vapor molecules,
$\lambda$ is the mean free path which may be estimated as follows.
We replace the toluene molecules by the impermeable
balls of the diameter $d\sim 1$ nm
(we found the estimate for the diameter by adding together
the lengths of the chemical bonds and taking into account
the molecule geometry). Therefore, the mean free path is
$
\lambda={1}/{(\pi\sqrt{2}nd^2)}=
{kT}/{(\pi\sqrt{2} P_0 d^2)} \approx 2.7\cdot 10^{-5}\mbox{\ cm.}
$
Here $n$ is the density of the toluene molecules in the toluene vapor.
If we substitute the obtained values into the Eq.~(\ref{diff-const}),
we get the estimate $D\sim 0.23$ cm$^2$/s.
This qualitative estimate is about in $1.5$ times larger than
the result obtained from comparison of our numerical
results and the experimental data in Sec~\ref{ResultsSec}.

Substituting the value of $D$ to the Eq.~(\ref{evap-time}),
we get the evaporation time of the spherical drop:
$t\approx 594 \mbox{\ s}$.

\subsection{Approximating a sessile drop surface with spherical caps}
\label{SessileSpherical}

Spherical caps offer a very simple model allowing for carrying out analytical calculations to the 
end and are thus widely used for the modeling of the 
droplet.
The equation for the surface is
\be
z(r)=\sqrt{\frac{r_0^2}{\sin^2\theta}-r^2}-\frac{r_0}{\tan\theta}.
\label{SphericalCap}
\ee

\begin{figure}[htb]
\caption{(Color online.) The sessile drop surface (red curve)
and three spherical caps (curves 1,2,3; see Table~\ref{s-table}).}
\label{4shapes}
\includegraphics[width=0.7\textwidth]{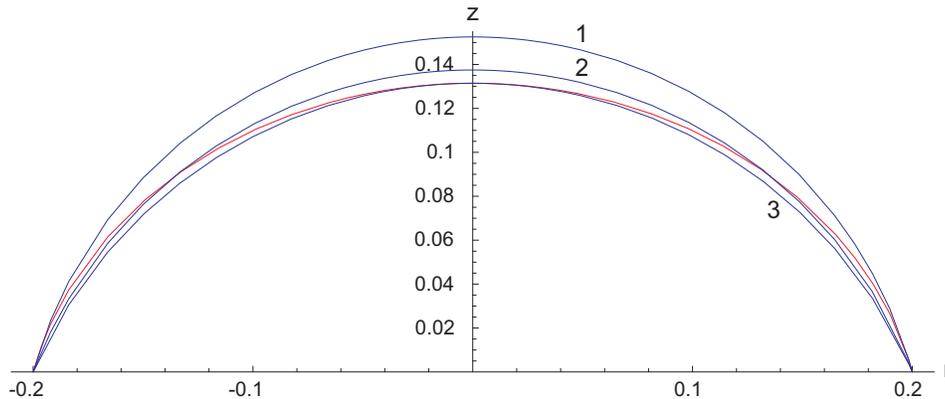}
\end{figure}

Figure~\ref{4shapes} shows surface of the sessile drop
and three approximating spherical caps with the parameters
listed in Table~\ref{s-table}. The radius of the substrate is
$r_0=0.2$ cm for all the drops. The first spherical surface
has the contact angle of the sessile drop.
The second spherical surface has the mass of the sessile drop.
The third spherical drop has the height of the sessile drop.

As seen from the Fig.~\ref{4shapes} and Table~\ref{s-table},
the characteristics $m$, $h$ and $\theta$ of all three approximating
spherical caps are quite close to those of the sessile drop.
This is not the case for local parameters such as
the local curvature of the sessile drop.
As follows from Eq.~(\ref{StaticLaplace}),
for the local curvature
to be approximately constant along the drop surface,
the condition $B_o=\rho ghr_0/(2\sigma\sin\theta) \ll 1$
has to be satisfied. Here $B_o$ is dimensionless number
which is analogous to the Bond number.

\vspace{0.5cm}
\begin{table}[htb]
\label{s-table}
\caption{Characteristics of the sessile drop and three spherical caps.}
\begin{tabular}{|c|c|c|c|c|}
\hline
Parameters & sessile drop & cap 1 & cap 2 & cap 3 \\
\hline
$m$, mg  &8.7 & 9.95625 & 8.7 & 8.22035 \\
\hline
$\theta$, radian &1.3032326& 1.3032326 & 1.20453564 & 1.16292249\\
\hline
$h$, cm  &0.13145179 & 0.152552 & 0.137494 & 0.13145179 \\
\hline
$\lambda(\theta)={(\pi-2\theta)}/{(2\pi-2\theta)}$
& 0.145545 & 0.145545 & 0.189081 & 0.206135 \\
\hline
curvature at top, cm$^{-1}$ & 8.0611 & 9.64418 & 9.33673 & 9.17966 \\
\hline
curvature at the contact line, cm$^{-1}$ &12.0727 & 9.64418 & 9.33673 & 9.17966 \\
\hline
\end{tabular}
\end{table}

\begin{figure}[htb]
\includegraphics[width=0.8\textwidth]{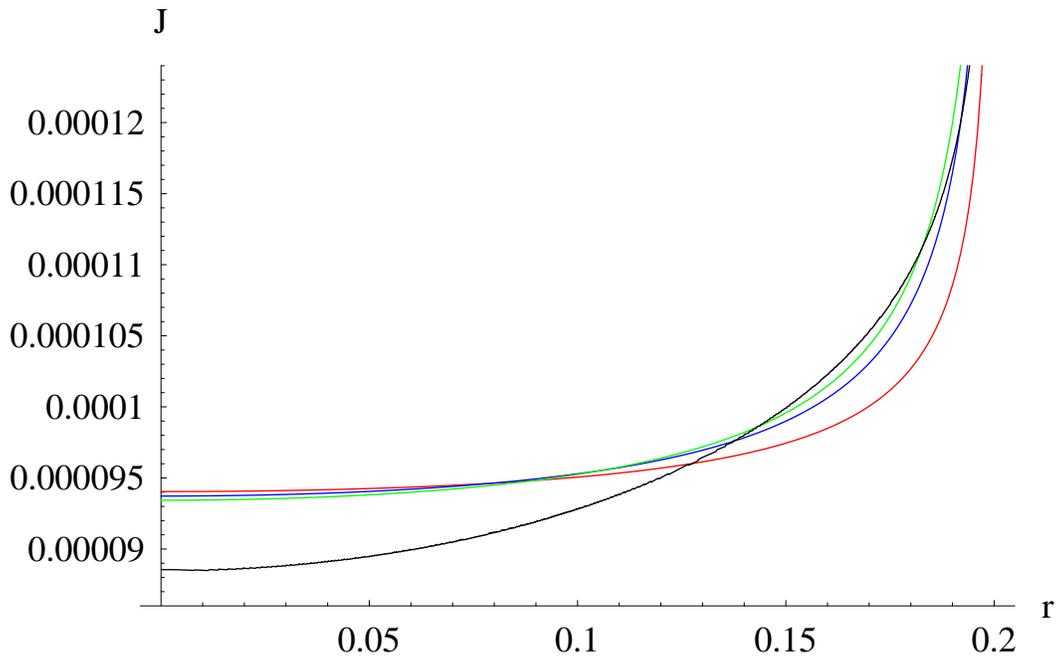}
\caption{
(Color online.)
Black curve -- $J(r)$ for the sessile drop with $m=8.7$ mg.
Red, blue and green curves -- $J(r)$ for the first,
the second and the third spherical drop corresponding
to notations of Table~\ref{s-table}.}
\label{sessile1}
\end{figure}

For our sessile drop $B_o\approx 0.4$; thus its curvature changes over
the surface by factor of $1.5$. Furthermore, one sees from
Fig.~\ref{sessile1} and Table~\ref{s-table}, that the assumption of the spherical cap shape
cannot provide an approximation for the local
evaporation flux density $J(r)$, which would have been accurate enough over a whole surface.
At the same time,
it is not surprising that the total evaporation rate,
integrated over the surface of the drop,
does not depend much on whether the drop is sessile or spherical.
With the values from Table~\ref{paramtable},
one may numerically integrate~(\ref{dmfromdt}) and obtain the
evaporation rates for the three spherical drops:
$dm_1/dt=20.52\ \ \mu\mbox{g/s}$,
$dm_2/dt=19.76\ \ \mu\mbox{g/s}$,
$dm_3/dt=19.47\ \ \mu\mbox{g/s}$.
The difference is within a few per cent.

\subsection{The Marangoni convection}
\label{MarangoniSec}

A fluid convection within a drop caused by the temperature dependent
surface tension which persists in the drop during the all stages of
an evaporation process is called the \textit{Marangoni convection}, and was
first observed in its classical form by B\'enard
in a process of formation of the characteristic hexagonal convection
patterns in flat fluid films. A theoretical
description of the Marangoni convection was developed by Pearson~\cite{Pearson}.
The Marangoni number $M_a$ (see Eq.~(\ref{Ma}) below)
characterizes a relative importance of surface tension
forces caused by the temperature variation and
viscous forces, leading to the Benard-Marangoni
instability if $M_a$ exceeds a critical value $M_c$.
For a flat fluid film, $M_c\approx 83$~\cite{Pearson}.
To estimate the Marangoni number for a drop we exploit a
formal similarity between a drop and a flat layer.
We extract the temperature difference between the substrate plane and the apex of
the drop from our simulation results as $\Delta T\approx 1$ K.
The drop height $h$, the thermal conductivity
$k$, the kinematic viscosity $\nu$, the specific heat $c_V$,
and the rate of the change of the surface tension with the temperature, $-d\sigma/dT$,
are taken from the Table~\ref{paramtable}.
Then the value of the Marangoni number at the beginning
of the drop evaporation process is

\be
M_a=\frac{-\frac{d\sigma}{dT} (c_V\Delta T) h}{\nu k} \approx 2800.
\label{Ma}
\ee

According to experimental data, a turbulent regime arises
for considerably larger values of the Marangoni number.
For example, in the water drop the turbulence takes place
for $M_a>22000$~\cite{Duan2005}. As the Reynolds number for toluene drop,
$R_e\approx 62$, is also comparatively small, the flow in our problem
is of a laminar character.
A competition between the buoyancy-induced flow and
the thermocapillary convection is characterized
by the dimensionless number
$B=\rho g h^2\beta/(7.14 \sigma')$~\cite{Pearson}.
In our case $B\approx 0.02\ll 1$, and
the buoyancy-driven convection is negligibly small.

For a flat liquid layer the height of the Marangoni cell
coincides with the layer thickness $h$, whereas the transverse size of the
cell is determined by the characteristic instability wavelength $\lambda$.
If the Marangoni number well exceeds the critical value,
then $\lambda=2\pi h/\alpha$ and $\alpha=\sqrt{M_a/8}$~\cite{Pearson}.

Our simulations show that near the surface a ``street" of vortices forms at
the initial stage of the convection (see Sec.~\ref{ResultsSec}).
The size of the vortices (i.e. the distance over which they extend into the bulk of the drop),
$h_d$, constitutes a noticeable fraction of the height $h$ of the drop.
With the assumption that the instability wavelength $\lambda$
and the Marangoni number can be described similarly to the case of
flat liquid layer, one can take the value $h_d$ for the layer thickness
and obtain $\lambda={2\pi\sqrt{hh_d}}/{\alpha}$.
Taking as an example the number of the near-surface vortices $N=9$ and $s_{\text max}=0.256$ cm, we find
$\lambda=s_{\text max}/N\approx 0.028$~cm and
\be
h_d=\left(\frac{\lambda\alpha}{2\pi h}\right)^2h\approx 0.4\, h.
\ee
This although crude estimate
turns out to be in a good agreement with our simulations
for the Marangoni convection
(see Figs.~\ref{karman-vect},\ref{karman-tv},\ref{vect1} and their
discussions).

\subsection{Hydrodynamic equations}
\label{EqSec}

The basic equations inside the drop are the Navier-Stokes equations,
the continuity equation for the incompressible fluid,
and the equation for thermal conduction
\begin{eqnarray}
\frac{\p\vvect}{\p t}+(\vvect\nabla)\vvect+
\frac1\rho\,{\mathrm{grad}\,p}
&=&\nu\,\Delta{\vvect},
\label{NavSt}
\\
\mydiv\vvect&=&0,
\label{ZeroDiv}
\\
\frac{\p T}{\p t}+\vvect\nabla T &=&
\kappa \Delta T.
\label{EqThermal}
\end{eqnarray}
Here $\Delta=\p^2/\p r^2+\p/r\p r+\p^2/\p z^2$,
$\nu=\eta/\rho$ is kinematic viscosity,
$\kappa=k/(\rho c_p)$ is thermal diffusivity.
Other terms in the thermal conduction equation
${\nu}/{(2c_p)}\left({\p v_i}/{\p x_k}+{\p v_k}/{\p x_i}\right)^2$,
where sum over $i$ and $k$ is assumed, are negligibly small.

Applying the ``$\rot$" operation  to the both sides of equation (\ref{NavSt}),
one excludes the pressure $p$ and obtains
\begin{equation}
\frac{\p}{\p t}\rot \vvect + (\vvect\nabla)\rot\vvect
-(\rot\vvect\cdot\nabla)\vvect =\nu \Delta\rot \vvect.
\end{equation}
Therefore,
\begin{equation}
\frac{\p}{\p t}\gamma(r,z)+(\vvect\nabla)\gamma(r,z)=
\nu \left(\Delta\gamma(r,z)-\frac{\gamma(r,z)}{r^2}\right),
\label{EqGamma}
\end{equation}
where the vorticity $\gamma$ is introduced by
\begin{equation}
\gamma(r,z)=\frac{\p v_r}{\p z}-\frac{\p v_z}{\p r};\qquad
\rot\vvect=\gamma(r,z)\mathbf{i}_\varphi.
\end{equation}
We define the stream function $\psi$, such that
\begin{equation}
\frac{\p\psi}{\p z}=rv_r\ , \qquad
\frac{\p\psi}{\p r}=-rv_z
\label{Psi}
\end{equation}
One has $\p^2\psi/(\p r\p z)=v_r+r(\p v_r/\p r)=-r(\p v_z/\p z)$,
therefore $v_r/r +{\p v_r}/{\p r}+{\p v_z}/{\p z}=0$, i.e.
the velocities obtained with Eqs.(\ref{Psi})
will automatically satisfy the requirement (\ref{ZeroDiv}).

Applying the Laplace operator to the stream function,
one obtains $\Delta\psi=r\gamma-2v_z$, which can be transformed
to a more convenient form $\tilde\Delta\psi=r\gamma$ with
the modified operator $\tilde\Delta$ that differs from
the Laplace operator by the sign of the term $\p/(r\p r)$:
\begin{equation}
\tilde\Delta\psi=
\frac{\p^2\psi}{\p r^2}-\frac1{r}\frac{\p\psi}{\p r}+
\frac{\p^2\psi}{\p z^2}=r\left(\frac{\p v_r}{\p z}-
\frac{\p v_z}{\p r}\right)=r\gamma.
\label{LaplacePsi}
\end{equation}

The method of the numerical solution is as follows:
at each step, inside the drop
\begin{enumerate}
\item We solve Eq.~(\ref{EqGamma}) with the proper boundary conditions
and find $\gamma(r,z)$;
\item We solve Eq.~(\ref{LaplacePsi}) and find $\psi(r,z)$.
Then we find velocities with Eqs.(\ref{Psi});
\item We solve Eq.~(\ref{EqThermal}) and obtain the new
boundary conditions for Eq.~(\ref{EqGamma}).
\end{enumerate}

Details of the numerical approach are given in Appendix~\ref{NumericalSec}.
The proper boundary conditions for quantities $\gamma$ and $\psi$,
satisfying Eqs.~(\ref{EqGamma}) and (\ref{LaplacePsi})
respectively, are derived in Appendix~\ref{BoundGammaPsi}.
They take the form
$\gamma=0$ for $r=0$; $\gamma=\p v_r/\p z$ for $z=0$;
$\gamma={d\sigma}/(\eta ds)+2v_\tau d\phi/ds$
on the surface of the drop; $\psi=0$ at all boundaries: at the
surface of the drop, at the axis of symmetry
of the drop ($r=0$), and at the bottom of the drop ($z=0$).

Here ${d\sigma}/{ds}=-\sigma'\p T/\p s$ is the derivative
of surface tension along the surface of the drop,
where the distribution of temperatures, which is found with
Eq.~(\ref{EqThermal}) is taken into account, and
\be
\sigma(T)=\sigma_0-\sigma'(T-T_0) \label{surface_tension},
\ee
is the experimental dependence of surface tension $\sigma$ on
temperature $T$.

The boundary conditions for the Eq.~(\ref{EqThermal}) are
$\p T/\p r=0$ for $r=0$; $T=T_0$ for $z=0$;
\be
{\p T}/{\p n}=-{Q_0(r)}/{k}
\label{ThermalBoundary}
\ee
on the surface of the drop.
Here $Q_0(r)=LJ(r)$ is the rate of heat loss per unit area
of the upper free surface, $J(r)$ is the local evaporation flux,
$T_0$ is the temperature of the substrate,
and $\mathbf{n}$ is a normal vector to the surface of the drop.

The relation $Q_0(r)=LJ(r)$ implies that the heat flow
from ambient air towards the drop surface is negligible.
This is the case if the temperature difference between
the drop surface and the air far from the drop
is less than $LDu_s/k$~\cite{Fuchs}.
This is well satisfied in the problem in question.

\section{Results}
\label{ResultsSec}

\subsection{Evaporation rates}

The evaporation rate of sessile drops is controlled mainly by
the vapor diffusion in the surrounding atmosphere~\cite{Deegan}.
Here we determined the diffusion coefficient of toluene vapor in air
by measuring the toluene evaporation rate.
The main conditions and parameters of the experiment are as follows.
The sessile drop of $10 \mu l$ of toluene is lying on a substrate
and evaporates to the atmosphere during $\approx 500$ s.
The contact line of the drop is pinned to the edge of the substrate.
The weight of the drop was measured every $10$ seconds.
As substrates, we used silicon wafers with a 100 nm thick
amorphous silicon nitride layer.

\begin{figure}[htb]
\caption{Drop evaporation rate $dm/dt$. Experimental data for the pure
toluene evaporation (open circles) and Au nanoparticle colloid (open
triangles). The solid line describes simulation results
obtained within the full numerical scheme for the vapor diffusion
(see also Sec.~\ref{DiffusionSec}).
The dotted line is obtained assuming the local evaporation flux to be
uniform over the drop surface and constant with time.
Inset: Mass variation at the end of evaporation.}
\label{fig-dmdt}
\includegraphics[angle=0,width=0.7\textwidth]{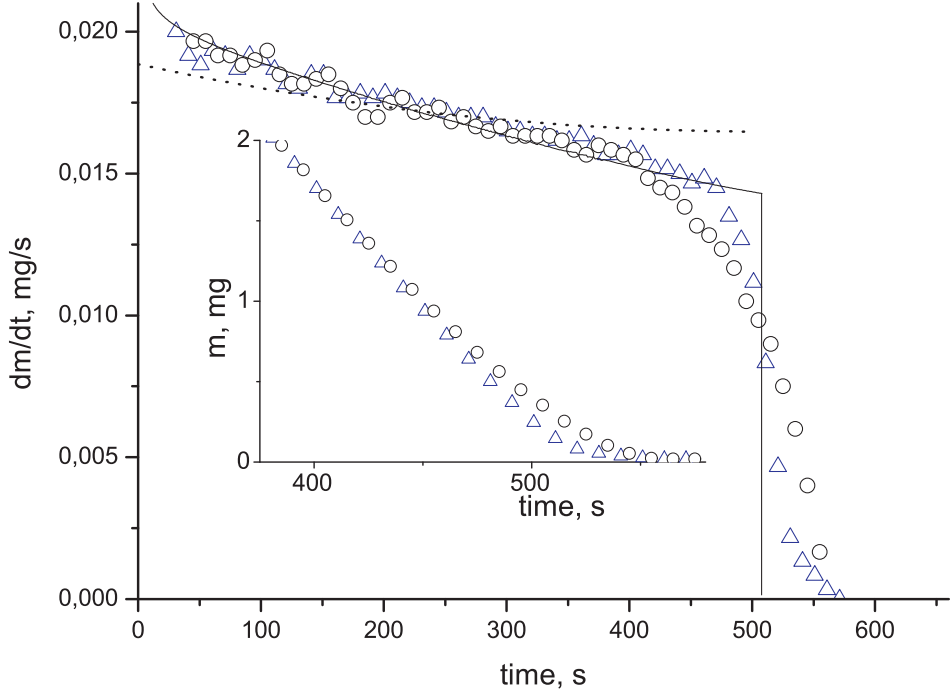}
\end{figure}

The results of the measurements of drop evaporation rates
are presented in Fig.\ref{fig-dmdt}.
Open circles are experimental values for the pure
toluene evaporation and open triangles are values for the
gold nanoparticles colloid.
The solid line is the simulation result obtained
in terms of Eq.~(\ref{dmdtEq}) within the full
numerical scheme. As seen in Fig.~\ref{fig-dmdt},
it agrees well with the experimental data.
Comparing the experimental data and the simulation results
permits us to find the diffusion coefficient of the vapor, $D$, the only
parameter controlling evaporation, as $D=0.1449$ cm$^2$/s.
It is worth noting that the rough estimate mentioned
in Sec.~\ref{RoughEstim} is larger by the factor of about 1.5.

One can divide the evaporation process
into three characteristic parts.
Our simulation describes the main part, i.e.
the quasi-stationary diffusion-limited evaporation,
which lasts for about $400$ seconds. While there
is a good agreement of the simulation and experimental
data during the diffusion-limited regime of evaporation,
this is not the case for the drying stage, i.e. the
final 50-100 seconds.
At the drying stage, the experimental evaporation rate
drops rapidly but not abruptly.
The main physical reason for such a behavior is the depinning
of the contact line. The interface area then shrinks over time.
The two distinct parts in the drying stage of evaporation can be
identified. The first part, that occurs in the $t{=}400{-}510$ time
window can be described as $dm/dt{\propto}(t_0{-}t)^\alpha$
that fits well the experimental data with
$t_0{=}550(2)$, $\alpha{=}0.41(2)$ for the evaporation
of the pure toluene and $t_0{=}522(3)$, $\alpha{=}0.28(8)$ for the colloid
evaporation. The last part of the drying stage is a pronounced
exponential decay of the evaporation rate
$dm/dt{\propto}\exp(-t/t_d)$, with the decay time $t_d{=}20.5(3)$
for the colloid evaporation.
The exponential behavior during this stage qualitatively
follows from the fact that change of the drop
surface area $A$ controls here the evaporation
rate evolution: ${dA}/{dt}\propto {dm}/{dt}\propto A$.
Inset in Fig.~\ref{fig-dmdt} represents the experimental mass variation
during the drying stage of the evaporation process.

\subsection{Street of vortices and the dynamics of vortex structure}

\begin{figure}[htb]
\caption{(Color online.)
The velocity field at $t=0.16$ {\rm s}.}
\label{karman-vect}
\includegraphics[width=0.6\columnwidth]{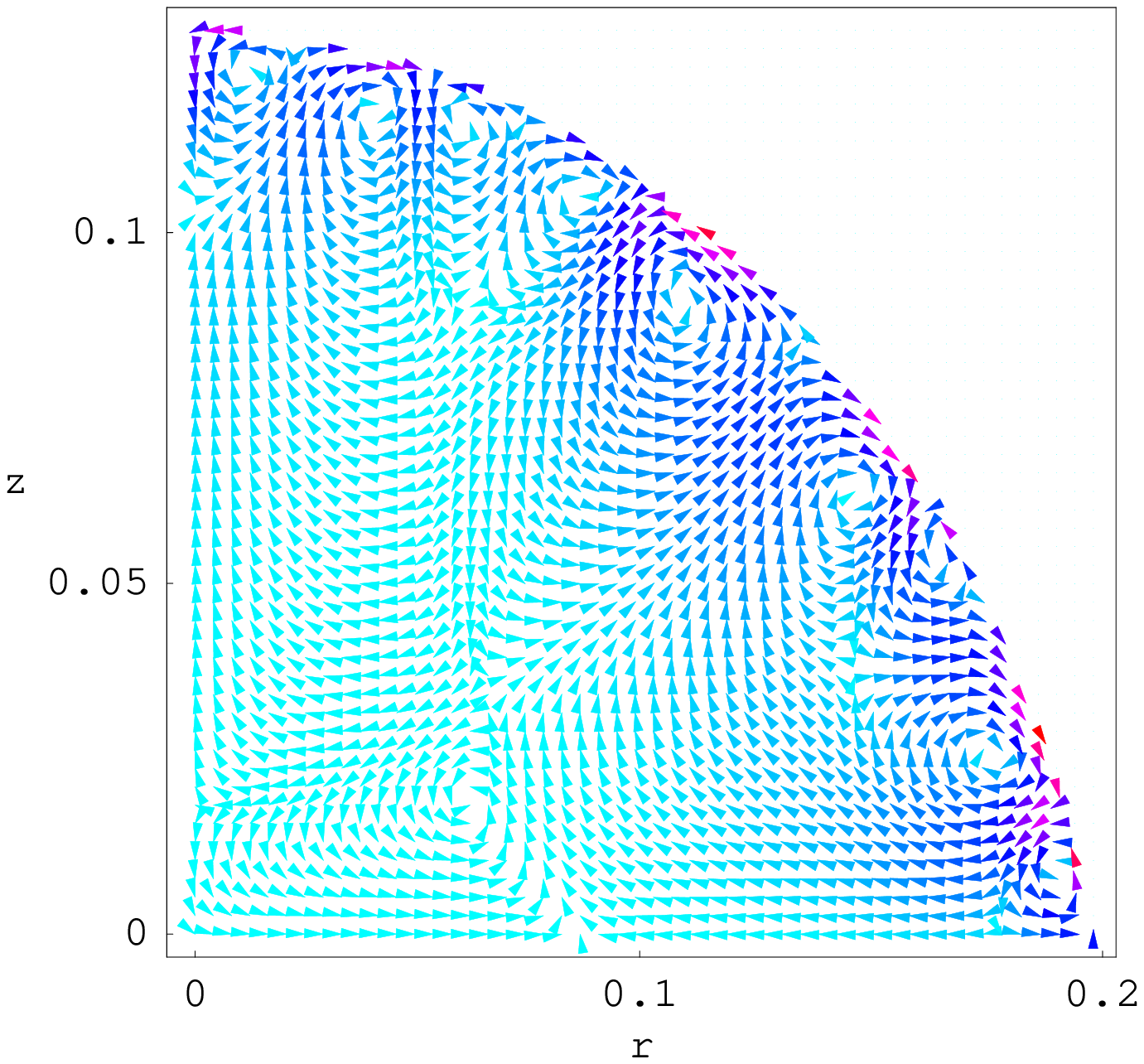}
\end{figure}

\begin{figure}[htb]
\caption{(Color online.)
Temperatures (curve T) and
velocities (curve v) along the drop surface as functions
of the radius at $t=0.16$ {\rm s}
(compare with Fig.~\ref{karman-vect}).
The velocity changes its sign at crossing points with
the horizontal axis.}
\hspace{0.047\columnwidth}
\includegraphics[width=0.631\columnwidth]{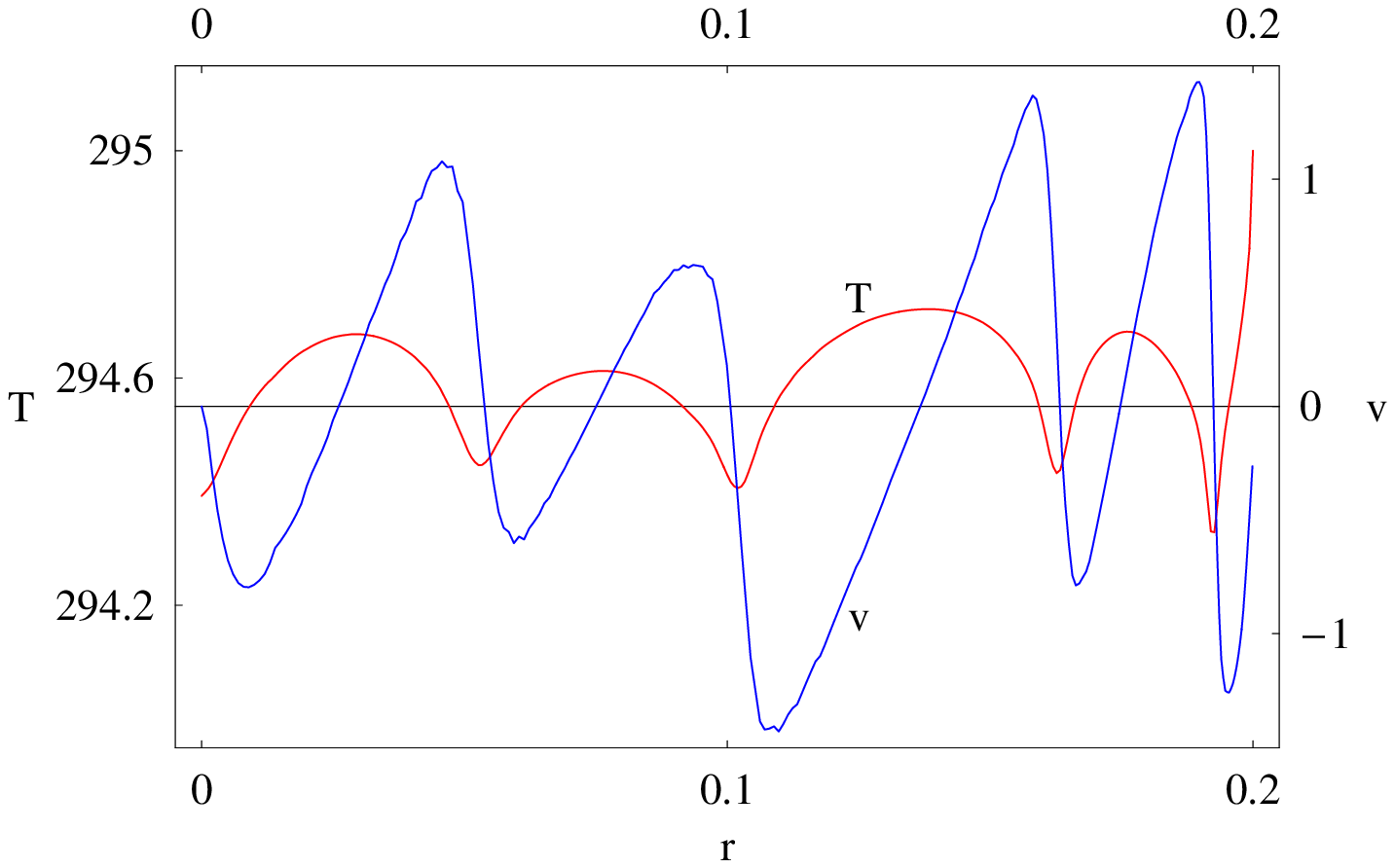}
\label{karman-tv}
\end{figure}

Our calculations demonstrate the presence of the
characteristic early regime in the dynamics of the Marangoni convection
in an evaporating sessile drop.
For various liquids and drop sizes, the vortices arise
near the surface of the drop.
For a toluene drop,
this regime quickly arises and
evolves up to $t\approx 0.3$ s.
This is quite a short time period
as compared with the total evaporation time $\approx 550$ s,
but it admits an experimental study.
The vortices grow,
the number of vortices decreases, and eventually
they evolve into the Marangoni convection
cells in the bulk of the drop. This can be seen
in Figs.~\ref{karman-vect},\ref{karman-tv},
where the vortex structure contains
six pairs of near-surface vortices and a
corner vortex, and the temperature
displays just six humps at the surface.
As was shown in Sec.~\ref{MarangoniSec},
this number of vortices in the drop
is controlled by the Marangoni cell size which is similar
to that given by Pearson for flat fluid layers.
The existence of near-surface vortices and the associated
humps in the profile of the surface temperature,
become more pronounced with
the decrease of the viscosity of the liquid.
There are no near-surface vortices when the viscosity
increases more than in four times as compared with
the toluene drop.

As seen in Fig.~\ref{karman-tv}, the extrema
of the surface temperature correspond to the
change of sign of the tangential component of
velocities at the surface. The reason for
this behavior is that the fluid flow moves
from the higher to the lower temperature regions
of the surface, because, according to (\ref{surface_tension}),
the surface tension decreases with increasing the temperature.
The flows result in a redistribution of the temperature
due to the convective heat transfer.
The number of the surface temperature humps and
the number of the near-surface vortices decrease during their evolution.

The initial conditions chosen when generating Fig.~\ref{karman-vect}
were: the vanishing velocities,
constant room temperature, and the vanishing vapor concentration.
For a description of a real experiment they have to be slightly modified to include the weak stochastic
distribution of the surface temperature and velocities.
We have carried out such calculations
using a small-scale stochastic initial conditions.
This has changed the particular behavior of the
fluid dynamics only on the initial stage of the process,
where a large number of small-scale surface vortices
arise. Then the surface vortex structure quickly evolves into
exactly the same one as we obtained for the basic
initial conditions. This result demonstrates the
generic character of the near-surface vortex regime
in Fig.~\ref{karman-vect} at the early
stage of the formation of the Marangoni convection.

An initial velocity field within the sessile drop
can also be strongly disturbed right after the drop has fallen
down on the substrate, or for some other reason.
We model such a situation by choosing random initial
conditions for the bulk velocity field, which are
in agreement with the continuity equation (\ref{ZeroDiv}).
We find that strong disturbances of the bulk
velocities can noticeably modify the initial
stage of the drop dynamics, but they do not
modify its main stage. For the initial random
velocities of the order of $5$ \rm{cm/s}
(which well exceeds the typical velocity in the vortex
$1$~cm/s),
the difference between the dynamics
of the disturbed and resting drops disappears
after $t\approx 0.5$ \rm{s}.
This verifies the stability of the large-scale drop dynamics
with respect to disturbances of the initial temperature
and velocity fields.

\begin{figure}[bht]
\caption{(Color online.) Left panel: The velocity distribution
at $t=0.5$ s. The stage of drop dynamics with three vortices
takes place from $t\approx 0.45$ s to $t\approx 2.0$ s.
Right panel: The velocity distribution at $t=30$ s.
A distribution with a single vortex takes place from
$t\approx 2.0$ s to $t\approx 250$ s.
} \label{vect2}
\includegraphics[width=0.49\columnwidth]{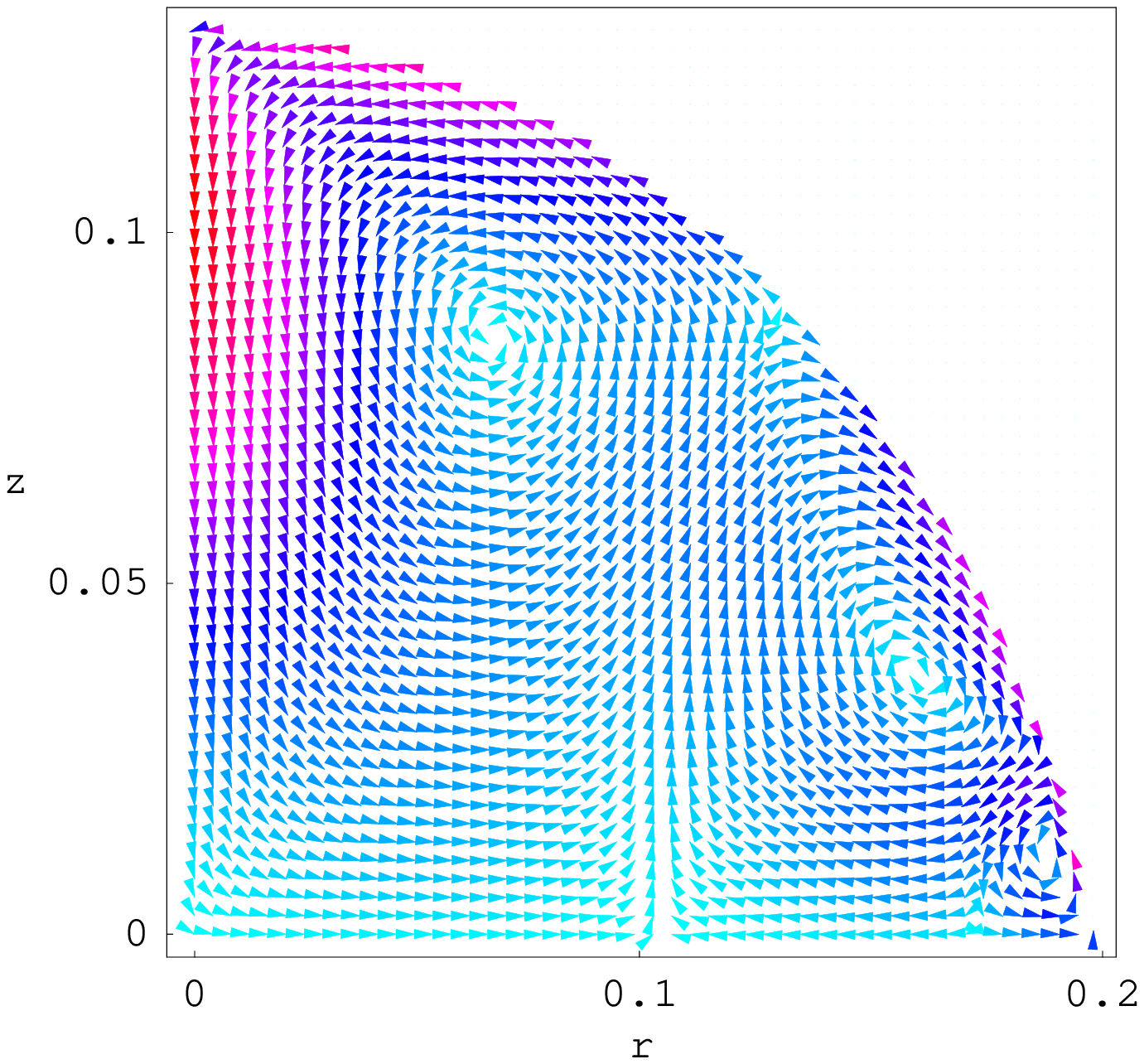}
\includegraphics[width=0.49\columnwidth]{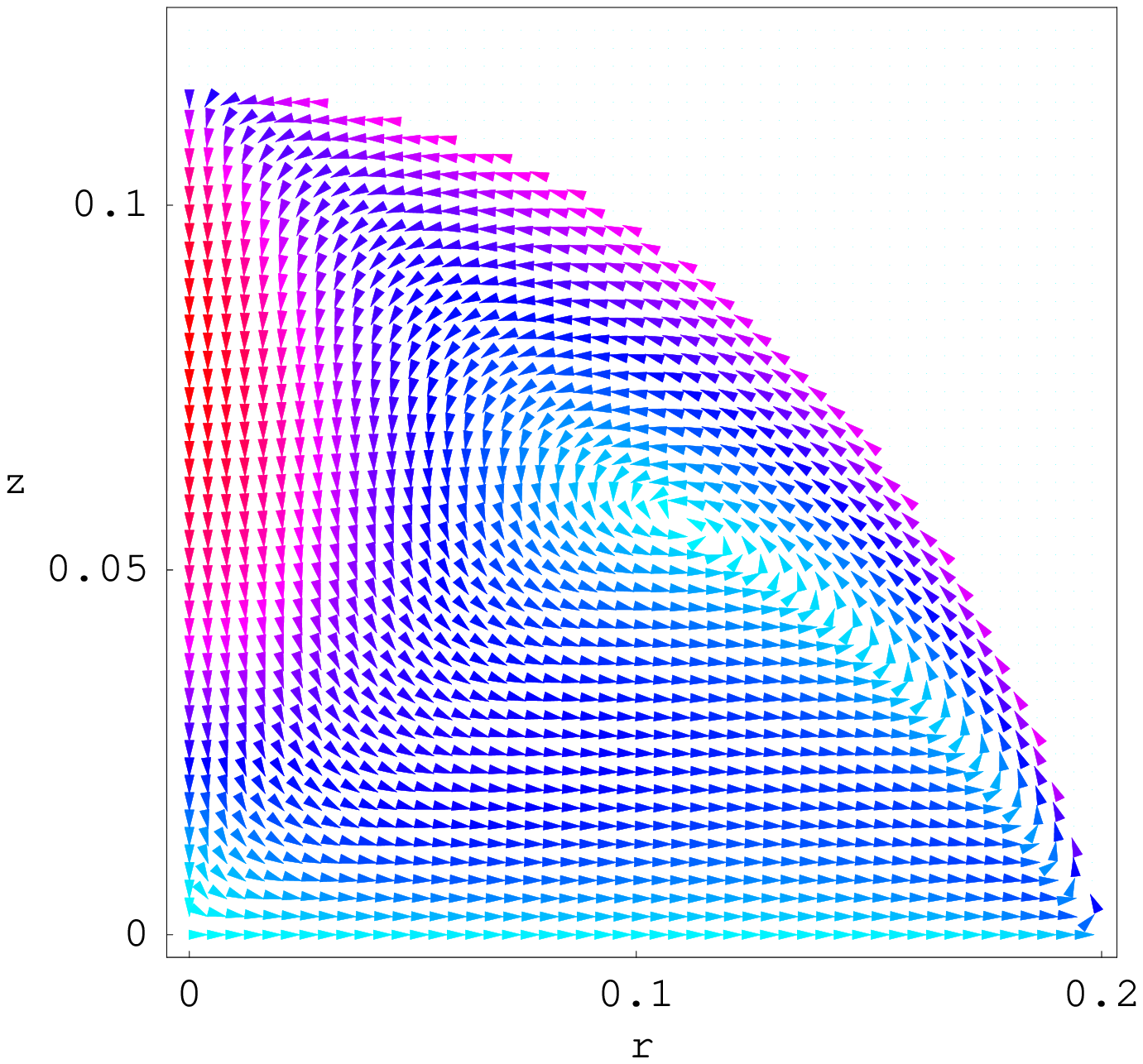}
\label{vect1}
\end{figure}

\begin{figure}[htb]
\caption{(Color online.) Distributions of temperature within drop for
the stages of drop dynamics with three vortices (left) and a single vortex
(right). The distributions are taken at $t=0.5$ s and at $t=30$ s.
correspondingly. Temperature scale shown at the right column.} \label{temp1}
\includegraphics[width=0.35\columnwidth,clip=true]{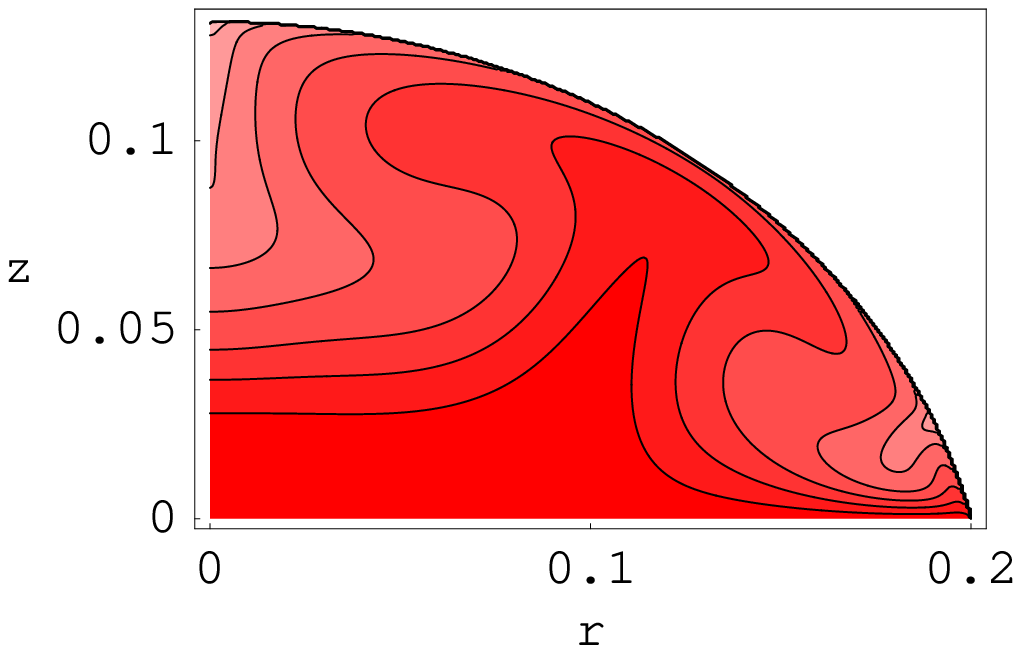}
\includegraphics[width=0.64\columnwidth,clip=true]{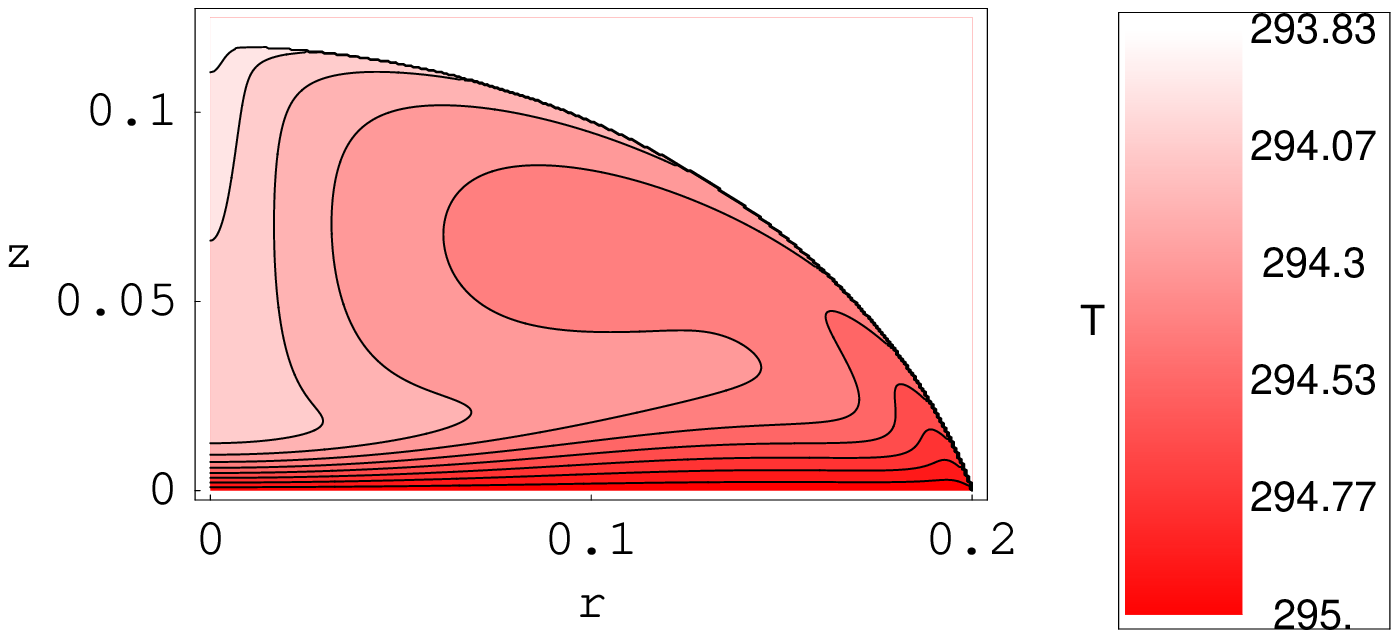}
\end{figure}

During the enlargement of the near-surface vortices,
their number decreases, and the convection involves
the bulk of the drop. As a result, for $t\approx 0.45$ s,
three bulk vortices control the velocity and temperature
fields in the drop, as seen in left panels in
Figs.~\ref{vect2},\ref{temp1}.
During the coexistence of three vortices, the corner vortex
starts growing at the expense of
the other two vortices, and eventually at $t\approx 2.0$ s
it occupies the whole drop volume.
A spatial dependence of the temperature along
the drop surface is nonmonotonic, if the
drop contains more than one vortex (see Fig.~\ref{temp1}).
Right panels in Figs.~\ref{vect1},\ref{temp1}
demonstrate how in the single-vortex regime
effects of Marangoni forces drive liquid
along the surface to the apex, where the fluid penetrates along
the symmetry axis in the depth of the drop.

The regime with the single vortex represents one of the
main stages of the dynamics of the evaporating sessile drop.
It lasts up to $t\approx 250$~s.
More than half of the drop mass evaporates during this period of time.
If the initial values of mass, height and contact angle of the drop are
$m=8.7$~mg, $h=0.1314$~cm, $\theta=1.2045$, then at the moment $t=250$~s
we find $m=4.0$~mg, $h=0.0685$~cm, $\theta=0.716$. In particular,
$h/(2r_0)\approx 0.17$, i.e. the drop shape is noticeably flattened.
The total time of the evaporation is $508$ s.

\begin{figure}[htb]
\caption{(Color online.)
Fluid pulsations at $t=263$ {\rm s.} (left panel) and at $t=281$ {\rm s}.
(right panel).}
\label{pulsations}
\includegraphics[width=0.49\columnwidth]{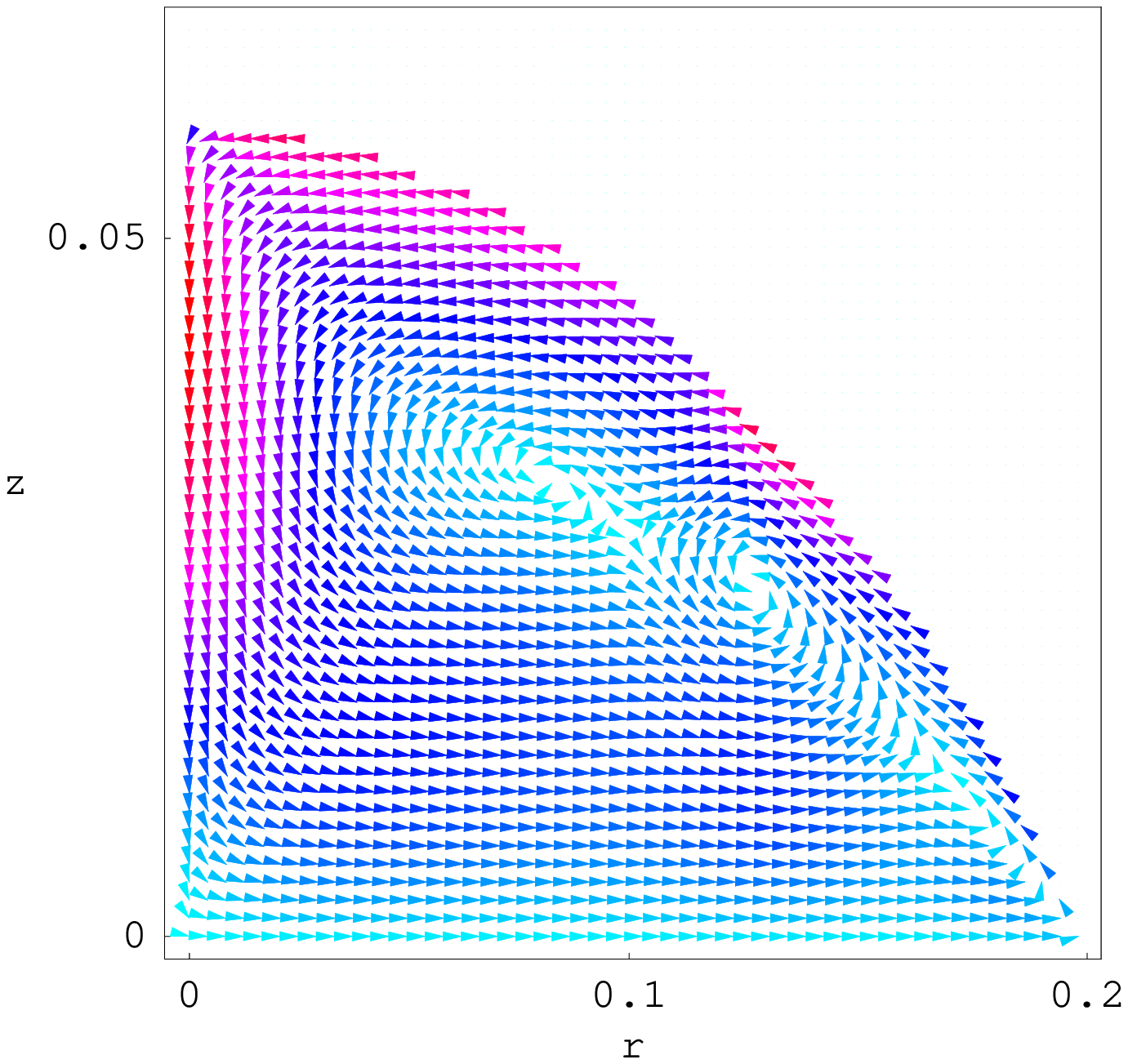}
\includegraphics[width=0.49\columnwidth]{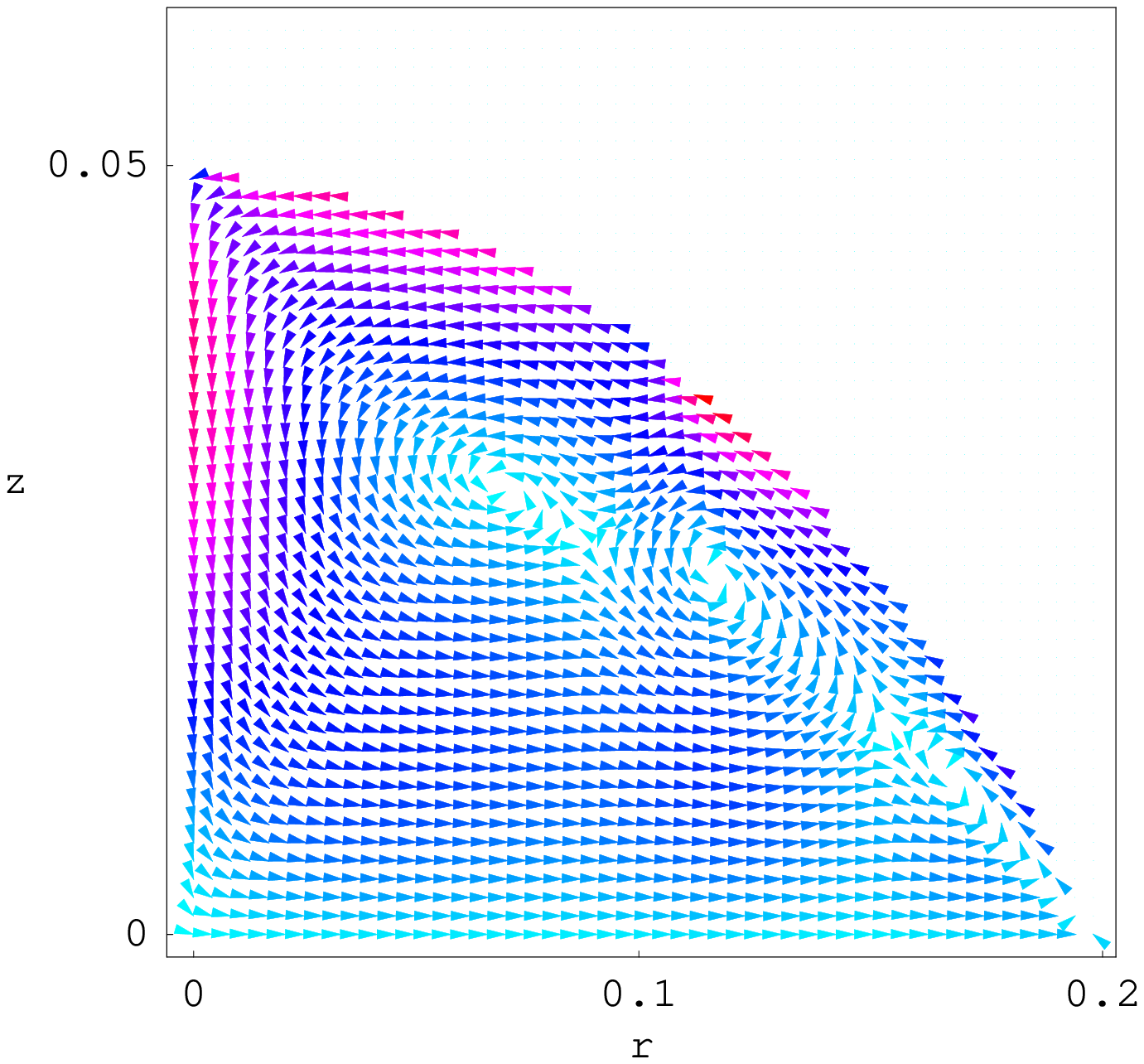}
\end{figure}

The quasistationary single-vortex state loses its stability at
$t\approx 250$ s and the vortex acquires a pronounced nonstationary
character. During this nonstationary regime, the fluid pulsations
take place. The characteristic frequency of the pulsations
corresponds to the circulation period $0.15$ s of a fluid element in
the original vortex. Initially the pulsations are concentrated near
the center of the original vortex. Then,
as shown in Fig.~\ref{pulsations}, the single-cell pulsating
state breaks into two-center (and later three-center) pulsating
structure. Eventually at $t\approx 300$ s a quasistationary state with
three vortices arises.

The numerical calculations of the fluid dynamics were tested with
several different mesh sizes. The respective results are
qualitatively identical and show reliable convergence
of the quantitative characteristics. For example,
the single-vortex regime was found to arise at $3.48$ s, $2.48$ s,
$2.2$ s, $2.06$ s, $2.01$ s for 100$\times$100, 150$\times$150,
200$\times$200, 250$\times$250, 300$\times$300 mesh elements
covering a half of the drop cross-section.

\subsection{Temperature profile}

If the thermal conductivity of a substrate is large compared to
that of the liquid, then the temperature can be maintained
practically constant at the substrate-fluid interface.
This is the case, in particular, for the silicon nitride substrates
used in experiments~\cite{Lin1,Lin2,Bigioni3}. The silicon nitride
is a material with high thermal conductivity, approximately in three
orders larger than for the toluene. For this reason,
the boundary condition for the temperature distribution
at the substrate can be reduced to the constant temperature.
Heat transfer between the substrate and the drop
plays an important role in establishing the temperature
profile in the drop. This also excludes a possibility for
the reversal of the Marangoni convection~\cite{Ristenpart},
taking place for substrates with relatively small thermal conductivity.

\begin{figure}[t]
\caption{Temperature profile along the surface as function of $r$,
in disregarding the effect of fluid flow on the thermal conduction.}
\label{fig-th}
\includegraphics[width=0.6\columnwidth]{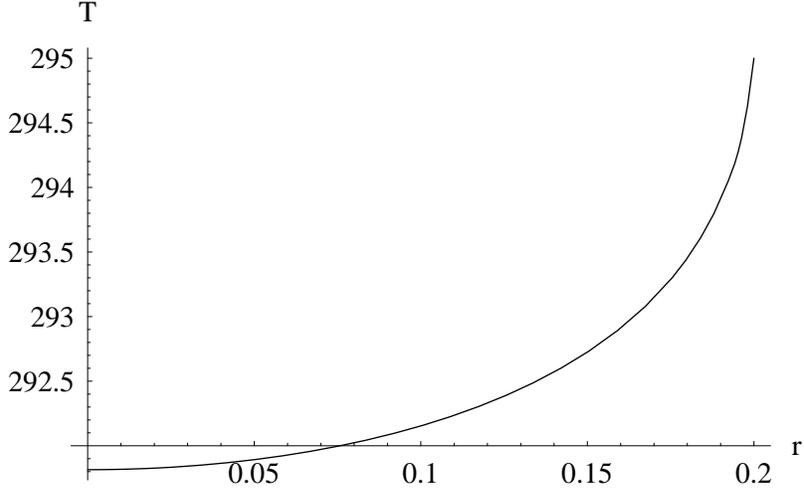}
\end{figure}

The characteristic scale of the temperature variation
on the drop surface can be easily estimated
if one disregards the effect of velocities on
the thermal conduction.
According to the evaporation rate data,
the heat loss per unit area of the drop near apex
is $Q_0=LJ(r=0)\approx 0.027$ W/cm$^2$ during the early
stage of the evaporation process. Therefore, as follows from
the boundary condition (\ref{ThermalBoundary}),
we obtain $|\p T/\p n|=22.88$ K/cm.
The calculations show that the temperature dependence
is almost linear on $z$ at $r=0$ in disregarding
the velocity term in thermal conduction.
This permits to obtain the temperature difference
between the substrate and the apex of the drop as
$\delta T \approx \left|{\p T}/{\p n}\right|_{r=0}h_0=3.0087 K$.
This estimate is in a good agreement with
the temperature profile along the surface
shown in Fig.~\ref{fig-th}, where the effect of
the velocity field on the temperature in
Eq.~(\ref{EqThermal}) is disregarded.
The temperature variation in Fig.~\ref{fig-th}
has larger amplitude and takes the monotonic form,
as compared with the respective curve
in Fig.~\ref{karman-tv}.
This demonstrates that the fluid flow noticeably
modifies the temperature variations in the drop.

Similar consideration for the evaporating water drop studied
in~\cite{HuLarsonMarangoni} results in $Q_0=0.093$~W/cm$^2$,
$|\p T/\p n|=15.49$~K/cm, $\delta T\approx 0.56$~K.
This estimation for the temperature difference exceeds in about $24$ times
the respective numerical results obtained in~\cite{HuLarsonMarangoni}
(see Figs.~1 and 2 there).
We consider this discrepancy
as an internal contradiction of~\cite{HuLarsonMarangoni} leading
to a significant underestimation of the velocity values in the water
drop.
Our numerical calculations confirm that $\delta T\approx 0.6$ K
for the water drop of~\cite{HuLarsonMarangoni}.
At the same time, our calculations exactly reproduce
the velocity field shown in Figs.~4 and 5 in~\cite{HuLarsonMarangoni},
if one takes a given surface temperature distribution
of~\cite{HuLarsonMarangoni}.
One should also note that
the role of convective term in thermal conduction equation
is substantially less for the water drop of~\cite{HuLarsonMarangoni}
due to a small size of the drop.

\subsection{Contact angles}

\begin{figure}[htb]
\caption{Contact angle of the drop during evaporation (solid line)
and angle of NCS domains orientation extracted from
experiment~\cite{Lin2} (solid circles) as function of time.}
\label{angle}
\includegraphics[width=0.5\columnwidth]{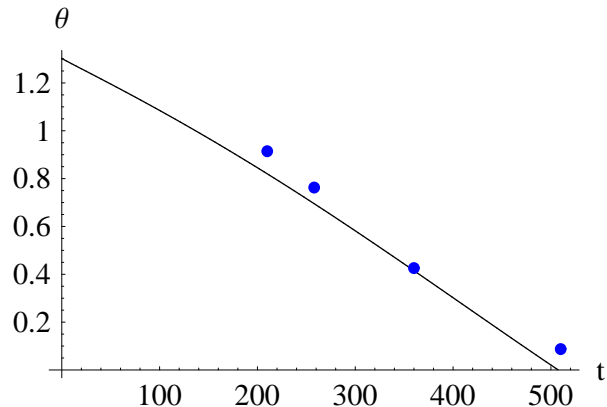}
\end{figure}

The evaporation process with the pinned contact line
is accompanied with an increase of the oblateness of the drop shape.
Figure~\ref{angle} displays our results for time dependence of the
contact angle of sessile drop.
Our results agree with the time-dependent contact angle
determined experimentally under the identical
initial drop parameters for the colloidal solution
of gold nanoparticles in the toluene drop~\cite{Lin2}.
In~\cite{Lin2} highly ordered nanoparticle islands
were found to form on the drop surface. For this reason
their orientation angle, which was
determined experimentally close to the contact line
with small angle X-ray scattering, coincides with
the drop contact angle.

\section{Discussion}
\label{DiscussionSec}

We have developed the approach for studying
the evaporation and fluid dynamics of a sessile drop of a capillary size
 and applied it for the description of the toluene drop evaporation.
In~\cite{Lin1,Lin2,Bigioni3}
the evaporating toluene drop containing colloidal
solution of gold nanoparticles was used for realizing
the self-assembly of nanocrystal monolayer.
During the evaporation process,
the nanoparticle islands were found to form on the drop
surface~\cite{Lin2}. For understanding
the island formation as well as nanoparticle segregation,
a study of the fluid and evaporation dynamics is necessary. In particular,
the convective flows described in this paper are important for driving the
nanoparticles to the contact line and the drop surface.

The numerical simulations were carried out and the system of
the diffusion equation for the vapor, the thermal conduction equation for
the temperature and the Navier-Stokes equations for the fluid flow in
the drop is solved.
The shape of the drop is controlled by the quasistationary Laplace
equation, which includes the effect of the gravitational forces.
Their role in forming a profile of a nonspherical sessile drop
is characterized by dimensionless number
$B_o=\rho ghr_0/(2\sigma\sin\theta)$, which is analogous to the Bond
number. Since in our case $B_o\approx 0.4$, effect of gravitational
forces, in general, should be taken into account.
We have found that deviations of
the sessile drop shape from the spherical cap are noticeable
in local drop characteristics such as the local curvature
and the evaporation flux density. At the
same time the integrated over the drop surface characteristics, like
the rate of the mass loss, are well described by the spherical cap
approximation.

The experimental and simulation results for the time dependent
drop evaporation rate agree well during the main
longest stage of the evaporation process. The calculated evolution of
the contact angle agrees with the results of~\cite{Lin2}.
In studying nonstationary diffusion equation, we found that the
time-dependent corrections to the stationary local evaporation
rates of the sessile drop do not vanish exponentially, but decay much
slowly $\propto 1/\sqrt{t}$ with time and, in general, can be
noticeable. This kind of behavior was established earlier only
for homogeneous diffusion from a surface.

Our solution demonstrates the presence of several time stages in the
evolution of the Marangoni convection in the drop.  The main
quasistationary fluid flow, containing only one vortex in the drop,
is formed during the longest period of the evaporation process.
Disregarding the time derivatives and
the dynamical behavior of the quantities
would result directly in a stationary state with one vortex
in the sessile drop for a given contact angle.
This kind of single-vortex solutions was obtained in previous
theoretical studies of the sessile drop evaporation
\cite{HuLarsonMarangoni,Girard}. Taking into account an explicit time
dependence of all quantities, permitted us to identify a sequence of
early dynamical stages of the Marangoni convection in the drop,
containing a street of near-surface vortices which then transforms
to the state with three bulk vortices.
The stability of the large-scale drop dynamics with respect to
perturbations of the initial conditions is identified.
Also, for a flattened evaporating drop we found under the conditions
in question a three-vortex state instead of the single-vortex one.
We establish an important role of inertial terms in the Navier-Stokes
equations and convective heat transfer terms in the
thermal conduction equation.

Our further plans in developing the present approach
include investigations of effects of surface surfactants and
nanoparticles, effects of a nonstationary dynamics of the
drop profile and a detailed study of the influence of
substrate properties on the drop dynamics.

Authors thank H.M.~Jaeger for fruitful and encouraging discussions.
This research was supported by the U.S. Department of Energy Office 
of Science under the contract DE-AC02-06CH11357 and by the Program 
of Russian Academy of Sciences. L.Yu.~Barash thanks the Dynasty 
Foundation for the partial support of his research.

\appendix
\section{Numerical methods}
\label{NumericalSec}
\subsection{A brief outline of the method}

The simultaneous calculation of the physical quantities in the drop can
be partitioned into several steps:

\begin{enumerate}
\item We apply to the diffusion equation (\ref{Diffusion})
the implicit finite difference method using irregular mesh
outside the drop and a variable time step. We use a boundary
interpolation in a vicinity of the drop surface.
For the boundary conditions we take $u=u_s$ on the drop surface,
$u=0$ far away from the drop, $\p u/\p r=0$ and $\p u/\p z=0$
on the axes $r=0$ and $z=0$ correspondingly.

\item Calculations of the stream function $\psi$ and
velocities $\vvect$ inside the drop are based on
Eqs.~(\ref{Psi}),~(\ref{LaplacePsi}).
The implicit finite difference method with a regular mesh
inside the drop is applied. We use a boundary interpolation
near the drop surface.
For the boundary conditions we take $\psi=0$ at all boundaries:
on the surface of the drop and on the axes $r=0$ and $z=0$.

\item We solve Eq.~(\ref{EqGamma}) to obtain
the vorticity $\gamma$ inside the drop.
The explicit finite difference method with a regular mesh
inside the drop is used. We use a boundary interpolation
close to the drop surface.
For the boundary conditions we take
$\gamma=0$ for $r=0$; $\gamma=\p v_r/\p z$ for $z=0$;
$\gamma={d\sigma}/(\eta ds)+2v_\tau d\phi/ds$
on the drop surface, where ${d\sigma}/{ds}=-\sigma'\p T/\p s$ is
the derivative of the surface tension along the drop surface
according to~(\ref{surface_tension}).

\item For calculating the temperature $T$ inside the drop,
the explicit finite difference method with a regular mesh
is applied to the thermal conduction equation~(\ref{EqThermal}).
We use a boundary interpolation in a vicinity of the drop surface.
The boundary conditions take the form $\p T/\p r=0$ for $r=0$;
$T=T_0$ for $z=0$; ${\p T}/{\p n}=-{Q_0(r)}/{k}=-LJ(r)/k$
on the drop surface.
Here $Q_0(r)$ is the rate of heat loss per unit area of the free
surface, $\mathbf{n}$ is a normal vector to the drop surface.

\item During the iterative procedure, the drop shape is recalculated
in accordance with the evaporative mass loss for the respective
time interval. For this purpose we use the
Runge-Kutta method for Eqs.(\ref{Cauchy1})-(\ref{Cauchy3}).

\end{enumerate}

\subsection{Explicit method for temperatures inside the drop}

For solving Eq.~(\ref{EqThermal}) the
following explicit finite difference method is used:
\begin{multline}
\frac{T_{ij}^{n+1}-T_{ij}^n}{h_t}
+v_{rij}\frac{T_{i+1,j}^n-T_{i-1,j}^n}{2h_r}
+v_{zij}\frac{T_{i,j+1}^n-T_{i,j-1}^n}{2h_z}
=\frac\kappa{h_r^2} (T_{i+1,j}^n-2T_{ij}^n+T_{i-1,j}^n)+\\
+\frac\kappa{h_z^2} (T_{i,j+1}^n-2T_{ij}^n+T_{i,j-1}^n)
+\frac\kappa{2ih_r^2} (T_{i+1,j}^n-T_{i-1,j}^n),
\end{multline}
which gives
\begin{multline}
T_{ij}^{n+1}=\alpha_z (T_{i,j+1}^n+T_{i,j-1}^n)
+\alpha_r (1+1/(2i)) T_{i+1,j}^n
+\alpha_r (1-1/(2i)) T_{i-1,j}^n
+(1-2\alpha_r-2\alpha_z) T_{ij}^n-\\
-\beta_r v_{rij} (T_{i+1,j}^n-T_{i-1,j}^n)
-\beta_z v_{zij} (T_{i,j+1}^n-T_{i,j-1}^n).
\label{TFDM}
\end{multline}
Here $T_{ij}^n$ is the temperature at the $n$-th time step,
$i,j$ are the coordinates on the regular mesh,
$h_r$,$h_z$ are steps along the respective axis,
$h_t$ is the time step,
\be
\alpha_r=\kappa h_t/h_r^2,\quad
\alpha_z=\kappa h_t/h_z^2,\quad
\beta_r=h_t/(2h_r),\quad
\beta_z=h_t/(2h_z).
\ee
For $r=0$ we have $\p T/\p r=0$, therefore for small $r$ one
has $T=\alpha+O(r^2)$, i.e. $\p^2 T/\p r^2=\p T/(r\p r)$.
Hence, instead of (\ref{TFDM}) we have the following formula:
\be
T_{0j}^{n+1}=\alpha_z (T_{0,j+1}^n+T_{0,j-1}^n)
+4 \alpha_r T_{1,j}^n
+(1-2\alpha_r-4\alpha_z) T_{0j}^n
-\beta_z v_{z0j} (T_{0,j+1}^n-T_{0,j-1}^n).
\ee
For $z=0$ we have $T_{i0}=T_0$.

\subsection{Boundary interpolation for temperatures inside the drop}

Let $G=\p T/\p n=-LJ(r)/k$, and $D(i,j)$ is the mesh point which is close
to the surface. The point $D$ is inside the drop and
at least one of its nearest neighbors has to be outside the drop.
In linear approximation $T=a+br+cz$
in a vicinity of the point $D(i,j)$. We denote the temperatures
at points $B(i-1,j)$ and $C(i,j-1)$ as
$u_B$ and $u_C$ correspondingly.
Then
\be
b\sin\phi+c\cos\phi = G,\qquad
a+b(i-1)h_r+cjh_z = u_B,\qquad
a+bih_r+c(j-1)h_z = u_C.
\ee
The solution of the set of equations is
\bea
a&=&\left(
h_r\cos\phi(iu_B-(i-1)u_C)+h_z\sin\phi(u_B+j(u_C-u_B))
-h_rh_zG(j+i-1) \right)/R\\
b&=&(h_zG+(u_C-u_B)\cos\phi)/R\\
c&=&(h_rG+(u_B-u_C)\sin\phi)/R\\
R&=&h_r\cos\phi+h_z\sin\phi.
\eea
The coefficients $a,b,c$ allow to calculate the temperature
at the point $D(i,j)$ using the formula $T_{i,j}=a+bh_ri+ch_zj$.
Also, the coefficients $a,b,c$ allow to find the temperature
at the surface of the drop near the point $D$.

\subsection{Alternating direction implicit method for
vapor density}

In order to cover quite a large region for the vapor density calculation
with Eq.~(\ref{Diffusion}), this is convenient to use irregular mesh.
We use the following mesh:
$r_i=i h_r$ for $i<n_r$,
$r_i=h_r(n_r+1.04^{i-n_r+79}-1.04^{79})$ for $i\ge n_r$,
$z_j=j h_z$ for $j<n_z$,
$z_j=h_z(n_z+1.04^{j-n_z+79}-1.04^{79})$ for $j\ge n_z$.
Here $n_r=n_z=700$, $n_r h_r=R$, $n_z h_z=h$.
This is the mesh with sufficiently small steps near
the drop surface and with exponentially increasing
steps, which are chosen in accordance with the
asymptotic decay of the vapor density
far away from the drop.

We denote the distances between the mesh point $(i,j)$
and its nearest neighbors as
$a=r_i-r_{i-1}$, $b=r_{i+1}-r_i$, $c=z_{j+1}-z_j$ and
$d=z_j-z_{j-1}$. Then the finite difference representations
for the second derivatives are

\bea
\hat\delta_r^2 u=\frac2{(a+b)ab}(au_{i+1,j}^n-(a+b)u_{ij}^n+
bu_{i-1,j}^n),\\
\hat\delta_z^2 u=\frac2{(c+d)cd}(du_{i,j+1}^n-(c+d)u_{ij}^n+
cu_{i,j-1}^n),
\eea
where $u_{ij}^n$ is the vapor density at mesh point $(i,j)$
for the $n$-th time step.

We apply the alternating direction method
to Eq.~(\ref{Diffusion}) with the above notations.
In the first part of the method one takes
$r$-derivative implicitly. Then the finite difference
representation of Eq.~(\ref{Diffusion}) is
\be
\frac{u_{ij}^{n+1/2}-u_{ij}^n}{Dh_t/2}=
\hat\delta_r^2 u_{ij}^{n+1/2}+\hat\delta_z^2 u_{ij}^{n}+
\frac{1}{r}\frac{u_{i+1,j}^{n+1/2}-u_{i-1,j}^{n+1/2}}{a+b}.
\ee
For given vapor density at time step $n$ it is
convenient to rewrite this expression as
\be
c_i'u_{i-1,j}^{n+1/2}+(d_i'-2/(Dh_t))u_{ij}^{n+1/2}
+e_i'u_{i+1,j}^{n+1/2}=
c_j''u_{i,j-1}^n+(d_j''-2/(Dh_t))u_{ij}^n+e_j''u_{i,j+1}^n.
\label{implicit1}
\ee
Here
\bea
c_i'=2/((a+b)a)-1/((a+b)r),& d_i'=-2/(ab),&
e_i'=2/((a+b)b)+1/((a+b)r),
\label{koeff1}
\\
c_j''=-2/((c+d)d),& d_j''=2/(cd),& e_j''=-2/((c+d)c),
\\
c_0'=0,& d_0'=-4/a^2,& e_0'=4/a^2,
\\
c_0''=0,& d_0''=2/c^2,& e_0''=-2/c^2.
\label{koeff4}
\eea
For each $j$ the tridiagonal matrix algorithm is used
to solve the set of equations (\ref{implicit1})
for the vapor density at the time step $n+1/2$.

In the second part of the method one takes
$z$-derivative implicitly and represents Eq.~(\ref{Diffusion}) as
\be
\frac{u_{ij}^{n+1}-u_{ij}^{n+1/2}}{Dh_t/2}=
\hat\delta_r^2 u_{ij}^{n+1/2}+\hat\delta_z^2 u_{ij}^{n+1}+
\frac{1}{r}\frac{u_{i+1,j}^{n+1/2}-u_{i-1,j}^{n+1/2}}{a+b}.
\ee
For given vapor density at time step $n+1/2$ this
expression takes the form
\be
c_j''u_{i,j-1}^{n+1}+(d_j''+2/(Dh_t))u_{ij}^{n+1}+e_j''u_{i,j+1}^{n+1}=
c_i'u_{i-1,j}^{n+1/2}+(d_i'+2/(Dh_t))u_{ij}^{n+1/2}+e_i'u_{i+1,j}^{n+1/2},
\label{implicit2}
\ee
where the coefficients are given in (\ref{koeff1})-(\ref{koeff4}).
For each $i$ the tridiagonal matrix algorithm is used
to solve the set of equations (\ref{implicit2})
for the vapor density at time step $n+1$.

\subsection{The boundary interpolation for vapor density}

Consider a mesh point $D(i,j)$ close to the surface.
The point $D$ is outside the drop and at least
one of its nearest neighbors is inside the drop.
In linear approximation $u(r,z)=a+br+cz$
in a vicinity of the point $D(i,j)$.
Consider mesh points $B(i+1,j)$, $C(i,j+1)$
and a point $A$ of the drop surface near $D$.
Then one has
\be
a+br_A+cz_A=u(A),\qquad
a+br_B+cz_B=u(B),\qquad
a+br_C+cz_C=u(C).
\ee
The solution of the set of equations is
\bea
\label{ukoeff1}
a&=&((r_Cz_B-r_Bz_C)u(A)+(r_Az_C-r_Cz_A)u(B)+(r_Bz_A-r_Az_B)u(C))/R,\\
b&=&(z_C(u(A)-u(B))+z_B(u(C)-u(A))+z_A(u(B)-u(C)))/R,\\
c&=&((r_B-r_C)u(A)+(r_C-r_A)u(B)+(r_A-r_B)u(C))/R,\\
R&=&(r_Cz_B-r_Bz_C)+(r_Az_C-r_Cz_A)+(r_Bz_A-r_Az_B).
\label{ukoeff4}
\eea
In the first part of the alternating direction method
the calculations of rows proceed towards smaller values
of $j$. For this reason one should consider here
$u(C)$ and $u(A)=u_s$ as given quantities, whereas
$u(B)$ and $u(D)$ are unknown. It is convenient under
these conditions to represent (\ref{ukoeff1})-(\ref{ukoeff4})
as
\bea
a=a_0+a_1u(B),\qquad
b=b_0+b_1u(B),\qquad
c=c_0+c_1u(B),
\eea
and obtain explicit expressions for $a_0,a_1,b_0,b_1,c_0,c_1$.
This results in linear relation between $u(D)$ and $u(B)$
\be
u(D)=a+br_D+cz_D=(a_0+b_0r_D+c_0z_D)+(a_1+b_1r_D+c_1z_D)u(B),
\ee
which can be transformed to the form
$c_i'u_{i-1,j}^{n+1/2}+d_i'u_{ij}^{n+1/2}+e_i'u_{i+1,j}^{n+1/2}=b_i'$.
This completes the set of equations (\ref{implicit1}) for the
tridiagonal matrix algorithm.
Here $c_i'=0$, $d_i'=1$, $e_i'=-(a_1+b_1r_D+c_1z_D)$,
$b_i'=a_0+b_0r_D+c_0z_D$.

To carry out the boundary interpolation
in the second part of the alternating direction
method, similar expressions can be derived to relate
$u(D)$ and $u(C)$.

The obtained values $b$ and $c$ also allow to find
the local evaporation rate at the surface point $A$:
\be
J(A)=D\left|\frac{\p u}{\p n}\right|=-D(b\sin\phi+c\cos\phi).
\label{ler}
\ee
The total drop evaporation rate is found from local values~(\ref{ler})
with numerical integration of~(\ref{dmdtEq}).

\subsection{Alternating direction implicit method for stream function}

The equation to be numerically solved is (compare with~(\ref{LaplacePsi}))
\be
\frac1\omega \frac{\p\psi}{\p t}=
\frac{\p^2\psi}{\p r^2}+\frac{\p^2\psi}{\p z^2}
-\frac{1}{r}\frac{\p\psi}{\p r}-r\gamma(r,z)
\label{LaplaceEqWithTime}
\ee
The finite difference representation for the second derivatives
on a regular mesh are
\be
\hat\delta_r^2 \psi_{ij}=(\psi_{i+1,j}-2\psi_{ij}+\psi_{i-1,j})/h_r^2,\qquad
\hat\delta_z^2 \psi_{ij}=(\psi_{i,j+1}-2\psi_{ij}+\psi_{i,j-1})/h_z^2.
\ee
Taking $r$-derivative implicitly in the first part of
the alternating direction method, we have for Eq.~(\ref{LaplaceEqWithTime})
\be
\frac{\psi_{ij}^{n+1/2}-\psi_{ij}^n}{\omega h_t/2}=
\hat\delta_r^2 \psi_{ij}^{n+1/2}+\hat\delta_z^2 \psi_{ij}^{n}-
\frac{1}{r}\frac{\psi_{i+1,j}^{n+1/2}-\psi_{i-1,j}^{n+1/2}}{2h_r}-r\gamma_{ij}.
\ee
For a given $\psi$ at time step $n$ it is convenient to rewrite
this expression as
\be
c_i'\psi_{i-1,j}^{n+1/2}+(d_i'-2/(\omega h_t))\psi_{ij}^{n+1/2}
+e_i'\psi_{i+1,j}^{n+1/2}=
c_j''\psi_{i,j-1}^n+(d_j''-2/(\omega h_t))\psi_{ij}^n+e_j''\psi_{i,j+1}^n+r\gamma_{ij},
\label{implicit1psi}
\ee
where
\bea
c_i'=(1+1/(2i))/h_r^2,& d_i'=-2/h_r^2&
e_i'=(1-1/(2i))/h_r^2
\label{koeff1psi}
\\
c_j''=-1/h_z^2,& d_j''=2/h_z^2,& e_j''=-1/h_z^2
\label{koeff4psi}
\eea
For each $j$ the tridiagonal matrix algorithm is used
to solve the set of equations (\ref{implicit1psi})
and obtain $\psi$ at time step $n+1/2$.

In the second part of the method one takes
$z$-derivative implicitly and represents Eq.~(\ref{LaplaceEqWithTime}) as
\be
\frac{\psi_{ij}^{n+1}-\psi_{ij}^{n+1/2}}{\omega h_t/2}=
\hat\delta_r^2 \psi_{ij}^{n+1/2}+\hat\delta_z^2 \psi_{ij}^{n+1}-
\frac{1}{r}\frac{\psi_{i+1,j}^{n+1/2}-\psi_{i-1,j}^{n+1/2}}{2h_r}-r\gamma_{ij}
\ee
For given $\psi$ at time step $n+1/2$ this expression takes the form
\be
c_j''\psi_{i,j-1}^{n+1}+(d_j''+2/(\omega h_t))\psi_{ij}^{n+1}+e_j''\psi_{i,j+1}^{n+1}=
c_i'\psi_{i-1,j}^{n+1/2}+(d_i'+2/(\omega h_t))\psi_{ij}^{n+1/2}+
e_i'\psi_{i+1,j}^{n+1/2}-r\gamma_{ij},
\label{implicit2psi}
\ee
where the coefficients are defined in~(\ref{koeff1psi}),~(\ref{koeff4psi}).

For each $i$ the tridiagonal matrix algorithm is used
to solve the set of equations (\ref{implicit2psi})
for the stream function $\psi$ at time step $n+1$.

\subsection{Boundary interpolation for stream function}

Consider a mesh point $D(i,j)$ close to the surface.
The point $D$ is inside the drop and
at least one of its nearest neighbors is outside of the drop.
In linear approximation $\psi(r,z)=a+br+cz$
in a vicinity of the point $D(i,j)$.
For mesh points $B(i-1,j)$, $C(i,j-1)$ and
a point $A$ of the drop surface near $D$ one gets
\be
a+br_A+cz_A = 0,\qquad
a+br_B+cz_B = \psi(B),\qquad
a+br_C+cz_C = \psi(C).
\label{psiabc}
\ee
We solve the set of equations~(\ref{psiabc}) and obtain
$a$, $b$, $c$ as functions of $\psi(B)$, $\psi(C)$.
The solution takes the form
$a=a_0+a_1\psi(B)$, $b=b_0+b_1\psi(B)$, $c=c_0+c_1\psi(B)$,
where $a_0$, $b_0$, $c_0$, $a_1$, $b_1$, $c_1$ are functions of $\psi(C)$.

In the first part of the alternating direction method
the calculations of rows inside the drop
proceed towards larger values of $j$.
For this reason one should consider here $\psi(C)$ as a given quantity,
whereas $\psi(B)$ and $\psi(D)$ are unknown.
In order to complete the set of equations~(\ref{implicit1psi}) for the
tridiagonal matrix algorithm we obtain the following relation
between $\psi(B)$ and $\psi(D)$
\be
\psi(D) = a+br_D+cz_D = (a_0+b_0r_D+c_0z_D)+(a_1+b_1r_D+c_1z_D)\psi(B).
\ee
Similarly, for the calculation of columns in the second part
of the alternating direction method, one obtains
$a=a_0+a_1\psi(C)$, $b=b_0+b_1\psi(C)$, $c=c_0+c_1\psi(C)$ and
\be
\psi(D) = a+br_D+cz_D = (a_0+b_0r_D+c_0z_D)+(a_1+b_1r_D+c_1z_D)\psi(C).
\ee

\subsection{Explicit method for vorticity}

For solving Eq.~(\ref{EqGamma}) the following explicit finite
difference method is used:
\begin{multline}
\frac{\gamma_{ij}^{n+1}-\gamma_{ij}^n}{h_t}
+v_{rij}\frac{\gamma_{i+1,j}^n-\gamma_{i-1,j}^n}{2h_r}
+v_{zij}\frac{\gamma_{i,j+1}^n-\gamma_{i,j-1}^n}{2h_z}
=\frac\nu{h_r^2} (\gamma_{i+1,j}^n-2\gamma_{ij}^n+\gamma_{i-1,j}^n)+
\\
+\frac\nu{h_z^2} (\gamma_{i,j+1}^n-2\gamma_{ij}^n+\gamma_{i,j-1}^n)
+\frac\nu{2ih_r^2} (\gamma_{i+1,j}^n-\gamma_{i-1,j}^n)
-\frac\nu{2i^2 h_r^2}\gamma_{ij},
\end{multline}
\begin{multline}
\gamma_{ij}^{n+1}=\alpha_z (\gamma_{i,j+1}^n+\gamma_{i,j-1}^n)
+\alpha_r (1+1/(2i)) \gamma_{i+1,j}^n
+\alpha_r (1-1/(2i)) \gamma_{i-1,j}^n+
\\
+(1-2\alpha_r-2\alpha_z-\alpha_r/(2i^2)) \gamma_{ij}^n
-\beta_r v_{rij} (\gamma_{i+1,j}^n-\gamma_{i-1,j}^n)
-\beta_z v_{zij} (\gamma_{i,j+1}^n-\gamma_{i,j-1}^n).
\label{GFDM}
\end{multline}
Here
\be
\alpha_r=\nu h_t/h_r^2,\quad
\alpha_z=\nu h_t/h_z^2,\quad
\beta_r=h_t/(2h_r),\quad
\beta_z=h_t/(2h_z).
\ee
\subsection{Boundary interpolation for vorticity}

Consider a mesh point $D(i,j)$ close to the surface.
The point $D$ is inside the drop and
at least one of its nearest neighbors has to be outside the drop.
In linear approximation $\gamma=a+br+cz$
in a vicinity of the point $D(i,j)$.
We denote the vorticity at points $B(i-1,j)$ and $C(i,j-1)$ as
$\gamma(B)$ and $\gamma(C)$ correspondingly.
The vorticity at the point $A$ on the drop surface near the point $D$ is
obtained as
$\gamma(A)={d\sigma}/{(\eta ds)}+2v_\tau{d\phi}/{ds}$
(see Appendix~\ref{BoundGammaPsi}).
Then
\bea
a+br_A+cz_A &=& \gamma(A),\\
a+b(i-1)h_r+cjh_z &=& \gamma(B),\\
a+bih_r+c(j-1)h_z &=& \gamma(C).
\eea
The solution of the set of equations is
\bea
a&=&(h_zr_A(\gamma(B)+(\gamma(C)-\gamma(B))j)+h_rz_A(\gamma(C)+(\gamma(B)-\gamma(C))i)-\gamma(A)h_rh_z(i+j-1))/R,\\
b&=&(z_A(\gamma(C)-\gamma(B))+h_z(\gamma(A)+\gamma(B)(j-1)-j\gamma(C)))/R,\\
c&=&(r_A(\gamma(B)-\gamma(C))+h_r(\gamma(A)-i\gamma(B)+(i-1)\gamma(C)))/R,\\
R&=&h_zr_A+h_rz_A-h_rh_z(i+j-1).
\eea
The coefficients $a,b,c$ allow to calculate $\gamma$
at the point $D(i,j)$ using the formula
$\gamma(D)=a+bh_ri+ch_zj$.

\subsection{Calculation of the drop surface}

For solving numerically Eqs.(\ref{Cauchy1}),~(\ref{Cauchy2})
for the drop surface in the form of $\mathbf{y}(s)=(r(s),\phi(s),z(s))^T$,
it is convenient to modify the initial conditions~(\ref{Cauchy3})
as follows:
\be
\mathbf{y}(0)=\left(R_0\delta,\delta,0\right)^T.
\label{Cauchy3a}
\ee
Here $r(s),\phi(s),z(s)$ are unknown functions,
$s\in \{0,s_{max}\}$, and $\delta=10^{-9}$
is introduced to fulfill
the condition $\phi(0)/r(0)=\delta/(R_0\delta)=1/R_0$.

Consider the set of points $s_k=hk$, where $k=0,1,\dots,K$ and $Kh=s_{max}$.
Eqs.~(\ref{Cauchy1}),~(\ref{Cauchy2}),~(\ref{Cauchy3a}) are solved
with the Runge-Kutta method:
\bea
\mathbf{y}_{k+1}&=&\mathbf{y}_k+\frac{h}{6}\left(\mathbf{p}_1+2\mathbf{p}_2
+2\mathbf{p}_3+\mathbf{p}_4\right),\\
\mathbf{p}_1&=&\mathbf{f}\left(s_k,\mathbf{y}_k\right),\\
\mathbf{p}_2&=&\mathbf{f}\left(s_k+\frac{h}2,\mathbf{y}_k+\frac{h}{2}\mathbf{p}_1\right),\\
\mathbf{p}_3&=&\mathbf{f}\left(s_k+\frac{h}2,\mathbf{y}_k+\frac{h}{2}\mathbf{p}_2\right),\\
\mathbf{p}_4&=&\mathbf{f}\left(s_k+h,\mathbf{y}_k+h\mathbf{p}_3\right).
\eea
To increase the accuracy further, an interpolation is used
for obtaining the values $s_{max}$ and $h$. Then,
the drop mass is calculated numerically with Eq.~(\ref{mass}),
and the evaporation rate is given by (\ref{dmdtEq}).

\section{Boundary conditions for vorticity and stream function}
\label{BoundGammaPsi}

For a derivation of the boundary condition at the
surface of the drop for the quantity $\gamma$, which
satisfies Eq.~(\ref{EqGamma}), we consider a vicinity of a point taken
at the surface of the drop. The $r$- and $z$- components of the velocity
can be expressed at the surface via the respective tangential
and normal components, lying in the $rz$-plane:
\bea
v_r &=& \ \ v_\tau\cos\phi+v_n\sin\phi,
\label{urutauun}\\
v_z &=& -v_\tau\sin\phi+v_n\cos\phi.
\label{uzutauun}
\eea
The angle $\phi$ between $r$- and $\tau$- projections of the velocity
depends, in general, on the coordinate $s$ along the surface.
Therefore,
\bea
\frac{\p v_r}{\p z} &=& \ \ \frac{\p v_\tau}{\p z}\cos\phi
- v_\tau\sin\phi\cdot\frac{d\phi}{dz}+\frac{\p v_n}{\p z}\sin\phi,
\label{durdz}\\
\frac{\p v_z}{\p r} &=& -\frac{\p v_\tau}{\p r}\sin\phi
- v_\tau\cos\phi\cdot\frac{d\phi}{dr}+\frac{\p v_n}{\p r}\cos\phi.
\label{duzdr}
\eea
It follows from Eqs. (\ref{urutauun}), (\ref{uzutauun}), (\ref{durdz}) and
(\ref{duzdr})
\bea
\gamma=\frac{\p v_r}{\p z}-\frac{\p v_z}{\p r}=
\left(\frac{\p v_\tau}{\p z}\cos\phi+\frac{\p v_\tau}{\p r}\sin\phi\right)
+v_\tau\left(\frac{d\phi}{dr}\cos\phi-\frac{d\phi}{dz}\sin\phi\right)
-\left(\frac{\p v_n}{\p r}\cos\phi-\frac{\p v_n}{\p z}\sin\phi\right)
=\nonumber\\
=\frac{\p v_\tau}{\p n}+v_\tau\frac{d\phi}{ds}-\frac{\p v_n}{\p s}=
\frac{\p v_\tau}{\p n}+v_\tau\frac{d\phi}{ds}.
\label{gamma-boundary1}
\eea
The boundary condition at the surface of the drop can be expressed
as~\cite{LL6}
\be
\left(p-p_v-\sigma\left(\frac1{R_1}+\frac1{R_2}\right)\right)n_i=
\eta\left(\frac{\p v_i}{\p x_k}+\frac{\p v_k}{\p x_i}\right)n_k-
\frac{\p\sigma}{\p x_i},
\ee
where the unit vector $\mathbf{n}$ is directed towards the vapor
along the normal to the surface.
Taking the tangential component of this equation,
we find
\bea
\frac{d\sigma}{ds}=\eta\left(\frac{\p v_i}{\p x_k}+
\frac{\p v_k}{\p x_i}\right)n_k\tau_i=\eta\left(\frac{\p v_i}{\p n}\tau_i+
\frac{\p v_k}{\p s}n_k \right)=\nonumber\\
=\eta\left(\frac{\p v_\tau}{\p n}
+\frac{\p v_n}{\p s}-v_k\frac{\p n_k}{\p s}\right)
=\eta\left(\frac{\p v_\tau}{\p n}-v_\tau\frac{d\phi}{ds}\right).
\label{vtau-boundary}
\eea
Here $\tau_i$ are the components of the unit vector $\pmb{\tau}$
tangential to the surface.

As follows from Eqs.(\ref{gamma-boundary1}) and (\ref{vtau-boundary}),
the boundary condition for Eq.~(\ref{EqGamma}), i. e. for the quantity
$\gamma(r,z)$ at the surface of the drop, takes the following form
\be
\gamma=\frac1\eta\frac{d\sigma}{ds}+2v_\tau\frac{d\phi}{ds} .
\ee
Consider now the boundary conditions for $\psi$, which satisfies
Eq.~(\ref{LaplacePsi}).
For $z=0$ we have ${\p\psi(r,z=0)}/{\p r}=-rv_z(r,z=0)=0$ and,
hence, $\psi(r,z=0)=const$. At the symmetry axis $r=0$ we have
${\p\psi(r=0,z)}/{\p z}=rv_r(r=0,z)=0$, therefore $\psi(r=0,z)=const$.
At last, on the outer surface of the drop we have
\be
\frac{\p\psi}{\p s}=\frac{\p\psi}{\p r}\cos\phi
-\frac{\p\psi}{\p z}\sin\phi=-rv_z\cos\phi-rv_r\sin\phi=-rv_n=0.
\ee
Therefore, $\psi=const$ on the surface of the drop, and, hence,
$\psi=const$ at all boundaries. Since only the derivatives of the
stream function $\psi$ enter expressions for physical quantities,
the particular value of the constant does not have any observable
consequences. For this reason, one can put $\psi=0$ throughout
the boundary.

\end{document}